\RequirePackage{fix-cm}
\documentclass[smallextended, natbib]{svjour3}       

\smartqed  
\usepackage[svgnames,table, dvipsnames]{xcolor}
  \definecolor{diffstart}{named}{Grey}
  \definecolor{diffincl}{named}{Green}
  \definecolor{diffrem}{named}{Red}
\emergencystretch=\maxdimen
\usepackage{graphicx}
\usepackage{balance}
\usepackage{enumitem}
\usepackage{tabulary}
\usepackage[subject={Todo},author={MG}]{pdfcomment}
\usepackage{float}
\usepackage{spreadtab}
\usepackage{courier}
\usepackage{multirow}
\usepackage[draft]{todonotes}   
\usepackage{amsfonts}
\usepackage{fancybox} 
\usepackage{cleveref}
\usepackage{color}
\usepackage{pifont}
\usepackage{booktabs}
\usepackage{xr}

\definecolor{light-gray}{gray}{0.90}
\usepackage{tcolorbox}
\tcbset{colback=blue!13}
\newcommand{\lstbg}[3][0pt]{{\fboxsep#1\colorbox{#2}{\strut #3}}}
\usepackage{listings}
  \thicklines
\usepackage[utf8]{inputenc}
\usepackage[T1]{fontenc}
\usepackage{amssymb,amsmath}
\newlength{\maxlen}
\newcommand{\maxleak}{491}

\newcommand{\maxnum}{100.00}
 
\newcommand{\percentwithmaxbar}[3][Emerald!50]{%
  \settowidth{\maxlen}{\maxnum}%
  \addtolength{\maxlen}{6\tabcolsep}%
  \FPeval\result{round(#2/#3:2)}%
  \FPeval\percent{round(#2*100/#3:1)}%
  \rlap{\color{Emerald!50}\hspace*{-.9\tabcolsep}\rule[-.05\ht\strutbox]{\result\maxlen}{.95\ht\strutbox}}%
  \makebox[\dimexpr\maxlen-\tabcolsep][r]{#2 (\percent\%)}%
}
\newcommand{\percent}[1]{\FPeval\result{round(#1*100/\maxleak:2)}{#1 (\result\%)}}

\newcommand\databar[3][Emerald!50]{%
  \FPeval\result{round(#3/\maxleak:4)}%
  \rlap{\textcolor{#1}{\hspace*{\dimexpr-\tabcolsep+.5\arrayrulewidth}%
        \rule[-0.3\ht\strutbox]{\result\maxlen}{2\ht\strutbox}}}%
  \makebox[\dimexpr\maxlen-2\tabcolsep+\arrayrulewidth][l]{#2}}
\def\header{Value column which is much wider}
\newcommand{\code}{\texttt}
\newcommand{\cmark}{\ding{51}}%
\newcommand{\xmark}{\ding{53}}%
\newcommand{\mohammad}[1]
{} 
\newcommand{\art}[1]
{\textcolor{green}{Artur: #1}} 
 \newcommand{\js}[1]
 {} 
\newcommand{\subRQ}[1]
	{\noindent \textbf{\emph{#1}}}
\newcommand{\category}[1] 
	{\noindent {\emph{{#1.}}}}
\newlength{\tablen}

\newcommand{\RQI}{What is distribution of leak types in studied projects?}
\newcommand{\RQII}{How are leak-related defects detected?}
\newcommand{\RQIII}{To what extent are the leak-inducing defects localized?}
\newcommand{\RQIV}{What are the most common root causes?}
\newcommand{\RQV}{What are the characteristics of the repair patches?}
\newcommand{\RQVI}{How complex are repairs of the leak-inducing defects?}

\newcommand{\resource}{233 }
\newcommand{\memory}{219 }

\begin{document}

\title{Memory and Resource Leak Defects and their Repairs in Java Projects}


\author{Mohammadreza Ghanavati \and \\
        Diego Costa \and
        Janos Seboek \and \\
        David Lo \and
        Artur Andrzejak
}


\institute{Mohammadreza Ghanavati \at
              \email{ghanavati@uni-heidelberg.de}           
           \and
           Diego Costa \at
              \email{diego.costa@informatik.uni-heidelberg.de}
           \and
           Janos Seboek \at
              \email{janos.seboek@stud.uni-heidelberg.de}
           \and
           David Lo \at
              \email{davidlo@smu.edu.sg}
           \and
           Artur Andrzejak\at
              \email{artur.andrzejak@informatik.uni-heidelberg.de}
}

\date{Received: date / Accepted: date}

\maketitle
\begin{abstract}
Despite huge software engineering efforts and programming language support, resource and memory leaks are still a troublesome issue, even in memory-managed languages such as Java. Understanding the properties of leak-inducing defects, how the leaks manifest, and how they are repaired is an essential prerequisite for designing better approaches for avoidance, diagnosis, and repair of leak-related bugs. 

We conduct a detailed empirical study on 491 issues from 15 large open-source Java projects. The study proposes taxonomies for the leak types, for the defects causing them, and for the repair actions. We investigate, under several aspects, the distributions within each taxonomy and the relationships between them. We find that manual code inspection and manual runtime detection are still the main methods for leak detection. We find that most of the errors manifest on error-free execution paths, and developers repair the leak defects in a shorter time than non-leak defects. We also identify 13 recurring code transformations in the repair patches. Based on our findings, we draw a variety of implications on how developers can avoid, detect, isolate and repair leak-related bugs. 


\end{abstract}
\section{Introduction}
\label{sec:intro}
Leaks are unreleased system resources or memory objects which are no longer used by an application. In memory-managed languages such as Java, C\#, or Go, a garbage collector handles memory management. Garbage collector uses object reachability to estimate object liveness. It disposes of any heap objects which are no longer reachable by a chain of references from the root objects. However, if an unused object is still reachable from other live objects, the garbage collector cannot reclaim the space. Aside from memory, finite \textit{system resources} such as file handles, threads, or database connections require explicit management specified in the code. It is the responsibility of the programmer to dispose of the acquired resource after using it, otherwise, a resource leak is likely.

Leak-related bugs are severe~\citep{bugStudyOpenSrc} and can finally result in performance degradation and program crash. Hence, they should be resolved at an early stage of development. However, due to their non-functional characteristics, leaks are likely to escape traditional testing processes and become first visible in a production environment. The root cause of a memory leak can differ from the allocation which exhausts the memory~\citep{CorkLeak2007}. Some leaks can only be triggered if an abnormal behavior occurs such as an exception or a race condition. These factors make leak diagnosis hard and error-prone.

Defects induced by memory and resource leaks are among the important problems for both researchers and practitioners. Microsoft engineers consider leak detection and localization as one of the top ten most significant challenges for software developers~\citep{LoPractitioners}. This problem is addressed by various researchers, tools, and programming languages. Many previous works targeted memory and resource leak diagnosis by leveraging static and dynamic analysis. Static analysis is used for leak detection via finding unclosed resources on different execution paths~\citep{CloserResourceLeak,trackerResourceLeak,HeapSafteyLeak,CheremResourceLeak,LeakChecker2014, weimarErrrorHandling}. The main challenge of static analysis is the lack of scalability and high rate of false positives. To mitigate this issue, researchers apply dynamic analysis techniques for leak diagnosis~\citep{SWATLeak2004, BellLeak2006, HoundLeak2009, CorkLeak2007, Leakbot2003, leakContainerXu2008,perfblower}.

Programming languages provide support for programmers to prevent occurrences of leak-inducing defects. For instance, Java 7 introduces a new language construct, called~\texttt{try}-with-resources\footnote{https://docs.oracle.com/javase/tutorial/essential/exceptions/tryResourceClose.html} to dispose of the objects that implement the~\emph{autoclosable} interface. Various open-source or proprietary tools (e.g., FindBugs\footnote{http://findbugs.sourceforge.net}, Infer\footnote{http://www.fbinfer.com}) also aim to help programmers to find the potential leaks in the software codebase. For example, FindBugs provides some rules\footnote{http://findbugs.sourceforge.net/bugDescriptions.html} to warn programmers about potential file descriptor leaks.

Despite the above-mentioned academic work, language enhancements, and tool supports, a number of challenges are still open. The impact of these efforts depends on whether they target prevalent or rare issue types, whether they can handle difficult cases, and whether their assumptions are realistic enough to be applicable in practice. Programming language enhancements such as \texttt{try}-with-resources or tool support such as FindBugs help to find only the resource leaks and no memory leaks. Many of the academics work are motivated by anecdotal evidence or by empirical data collected only from small sets of defects. For example, \citet{leakContainerXu2008} propose a method for detecting memory leaks caused by obsolete references from within object containers but provide only limited evidence that this is a frequent cause of leak-related bugs in real-world applications. As another example, Leakbot~\citep{Leakbot2003} introduces multiple sophisticated object filtering methods based on observations derived from only five large Java commercial applications. 

A systematic empirical study of a large sample of leak-related defects from real-world applications can help both researchers and practitioners to have a better understanding of the current challenges on leak diagnosis. We believe such a study can be beneficial in the following directions:

\subRQ{Benefit 1.} A representative study can characterize the current approaches for leak diagnosis used in practice. This can guide researchers to find limitations of leak detection approaches and motivate further improvements. The results would provide a comprehensive basis for the design and evaluation of new solutions.

\subRQ{Benefit 2.} It helps programmers to avoid mistakes made by other programmers and shows some of the best practices for leak diagnosis. 

\subRQ{Benefit 3.} It can be used as a verification for the assumptions used in previous work. For example, it is interesting to verify empirically whether there is a large number of leaks caused by collection mismanagement in real-world applications. The positive answer to this could confirm the assumption of \citet{leakContainerXu2008} on memory leak detection.

To the best of our knowledge, the research body of empirical studies on resource and memory leak-related defects is relatively thin in comparison with the large body of studies about other bug types (e.g., semantic or performance bugs). The existing studies~\citep{fumio,bugStudyOpenSrc} provide only limited information about characteristics of detection types, root causes, and repair actions of leak defects. To fill this gap, we conduct a detailed empirical study on 491 real-world memory and resource leak defects gathered from 15 large, open-source Java applications. 

We manually study the collected issues and their properties: leak types, detection types, common root causes, repair actions, and complexity of fix patches. Based on our findings, we draw several implications on how to improve avoidance, detection, localization, and repair of leak defects. In particular, this study tries to answer the following research questions:

\begin{itemize}
\item[\textbf{.}] \textbf{RQ1.} \RQI
\item[\textbf{.}] \textbf{RQ2.} \RQII
\item[\textbf{.}] \textbf{RQ3.} \RQIII
\item[\textbf{.}] \textbf{RQ4.} \RQIV
\item[\textbf{.}] \textbf{RQ5.} \RQV
\item[\textbf{.}] \textbf{RQ6.} \RQVI
\end{itemize}

\noindent The preliminary idea of this work is presented in a two-page short paper in ICSE 2018~\citep{PosterICSE2018}. This work provides the following contributions:

\subRQ{Characterization study.} We conduct an empirical study on 491 bugs from 15 mature, large Java applications. To the best of our knowledge, this is the first work which studies characteristics of leak-related bugs from real-world applications in a comprehensive way while using a large set of issues from diverse open-source applications. 

\subRQ{Taxonomies.}
We propose taxonomies for leak types~(\Cref{RQ100}), detection types and methods~(\Cref{RQ200}), root causes~(\Cref{RQ400}), and repair actions~(\Cref{RQ500}).

\subRQ{Analysis.}
We investigate the distributions of leaks across the categories within each taxonomy and the relation between the taxonomies.
Our findings show that source code analysis and resource monitoring are the main techniques to detect leaks. Our analysis using a state-of-the-art resource leak detection tool (i.e., Infer) highlights that the static analysis tools require further improvement to detect different leak types in practice. We find that 76\% of the leaks are triggered during the error-free execution paths. We identify 13 recurring code transformations in the repair patches. We also show that developers resolved the studied issues in about 6 days on the median.

\subRQ{Implications.}
We use our findings to draw a variety of implications on the leak prevention and diagnosis for both researchers and practitioners~(\Cref{sec:implications}).

\subRQ{Replicability.} To make our study replicable and reusable for the community, we make the dataset and the results available online\footnote{https://github.com/heiqs/leak\_study}.

This paper is organized as follows.~\Cref{sec:background} provides a short background about leak definition and issues in the bug tracking systems.~\Cref{sec:methodology} describes the design of our empirical study. In~\Cref{sec:results}, we present the answers to the research questions. In~\Cref{sec:implications}, we present the implications drawn from our observations and findings.~\Cref{sec:threats} discusses potential threats to the validity of our study.~\Cref{sec:relatedWork} surveys related work. Finally,~\Cref{sec:conclusion} concludes the paper.

\section{Background}
\label{sec:background}
\subsection{Leak definition}
Leaks occur due to mismanagement of memory or finite systems resources. In this section, we briefly explain these two types.

\subRQ{Memory leak.}
Contrary to the unmanaged languages such as C or C++ in which programmer is responsible for freeing the memory, in memory-managed languages such as Java or C\#, a garbage collector reclaims the space. A programmer can rely on the garbage collector to release references due to dangling pointers or lost pointers. However, if the references to the unused objects are present in the running process, they cannot be garbage-collected. As a sequence, a memory leak might be triggered. In other words, a~\emph{memory leak} in Java occurs when a process maintains unnecessary references to some unused objects. 

\subRQ{Resource leak.}
In Java, finite system resources like connections, threads, or file handles are wrapped in special~\emph{handle objects}. Programmer accesses such a resource by normal object allocation. However, in contrast to memory management, the developer should dispose of a system resource by making an explicit call to the disposal method of the handle object (or by ensuring that a thread has stopped). Besides this, all unnecessary references to such objects should be removed to prevent the potential memory leak. Hence, a~\emph{resource leak} occurs when the programmer forgets to call the respective close method for a finished handle object. Similar to the memory leak, a resource leak gradually depletes system resources which degrades performance and can lead to a failure.

In this paper, we use the term~\emph{leak} for both memory and resource leaks. We also occasionally use the term~\emph{disposing of an object} for either closing a resource or releasing (deallocating) memory (in Java, by removing all references to an object). 
\begin{figure}[t]
\centering
\scalebox{0.6}{
\includegraphics{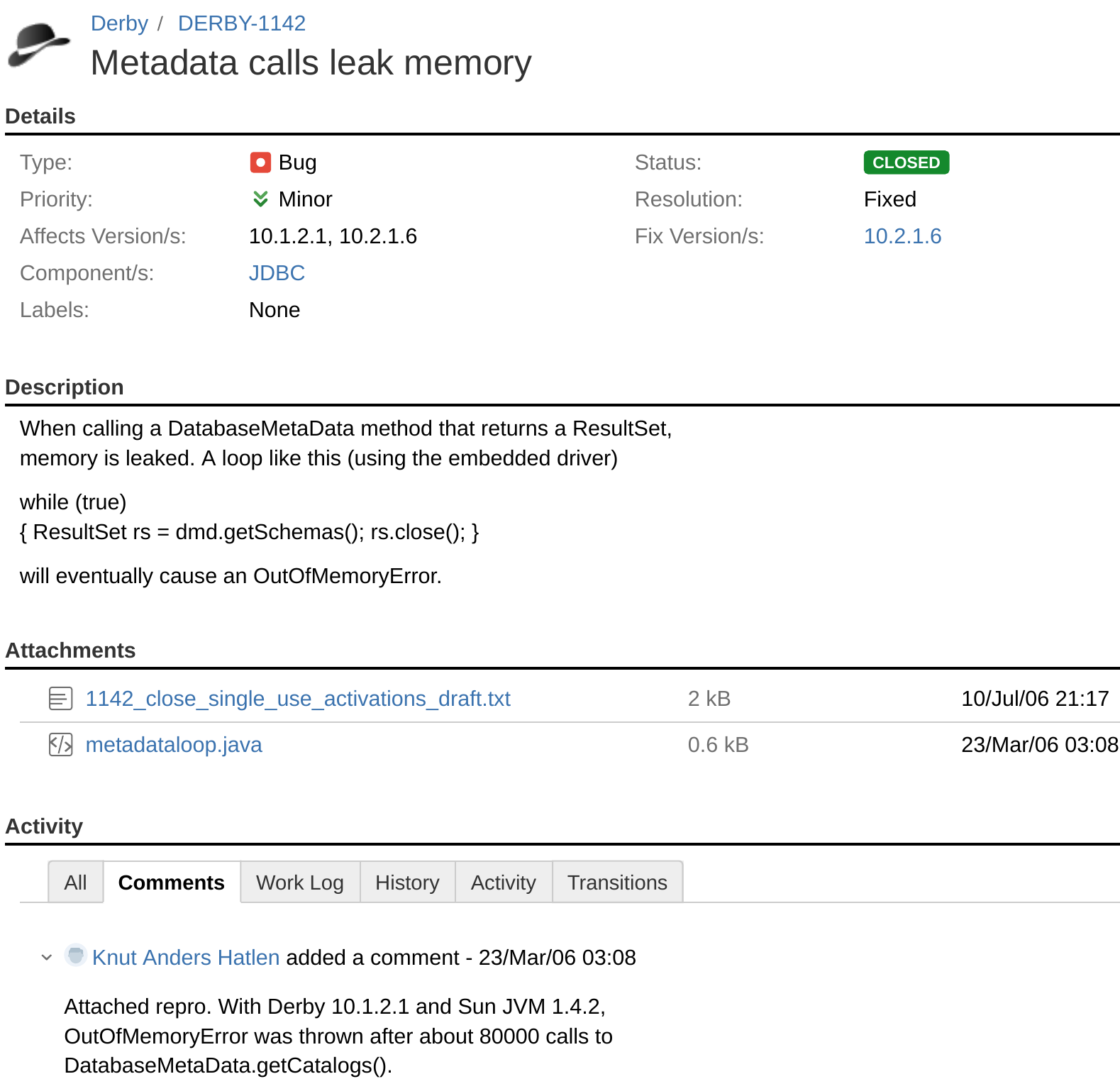}}	
\caption{An issue report from JIRA.}
\label{fig:issue}
\end{figure}

\subsection{Issue Report}
Modern projects often use an Issue Tracking System (ITS) to collect the issues reported by users, developers, or software quality teams. An issue typically corresponds to a bug report or a feature request. Bugzilla\footnote{https://www.bugzilla.org/}, JIRA\footnote{https://issues.apache.org/jira/projects/}, and GitHub issue tracker\footnote{https://github.com/} are examples of ITS systems. Each issue report in the bug tracker is identified with a unique identifier. For example, in JIRA, this is a combination of the project name and a number (e.g., SOLR-1042). In GitHub, an identifier is a number with a preceding hashtag (e.g., issue \#1865 in RxJava project). An issue report in Jira contains a variety of information such as title, description, comments, and links to the related fix patches. It also contains metadata information such as type, status, priority, resolution, and associated timestamps (e.g., created or resolved timestamps).~\Cref{fig:issue} shows a snippet of an issue report from Jira. All the information provided in issue reports makes the issue tracker a rich environment to get more insights on bugs and their corresponding repairs.

\section{Empirical Study Design}
\label{sec:methodology}
\begin{figure}[t]
\scalebox{0.55}{
\includegraphics{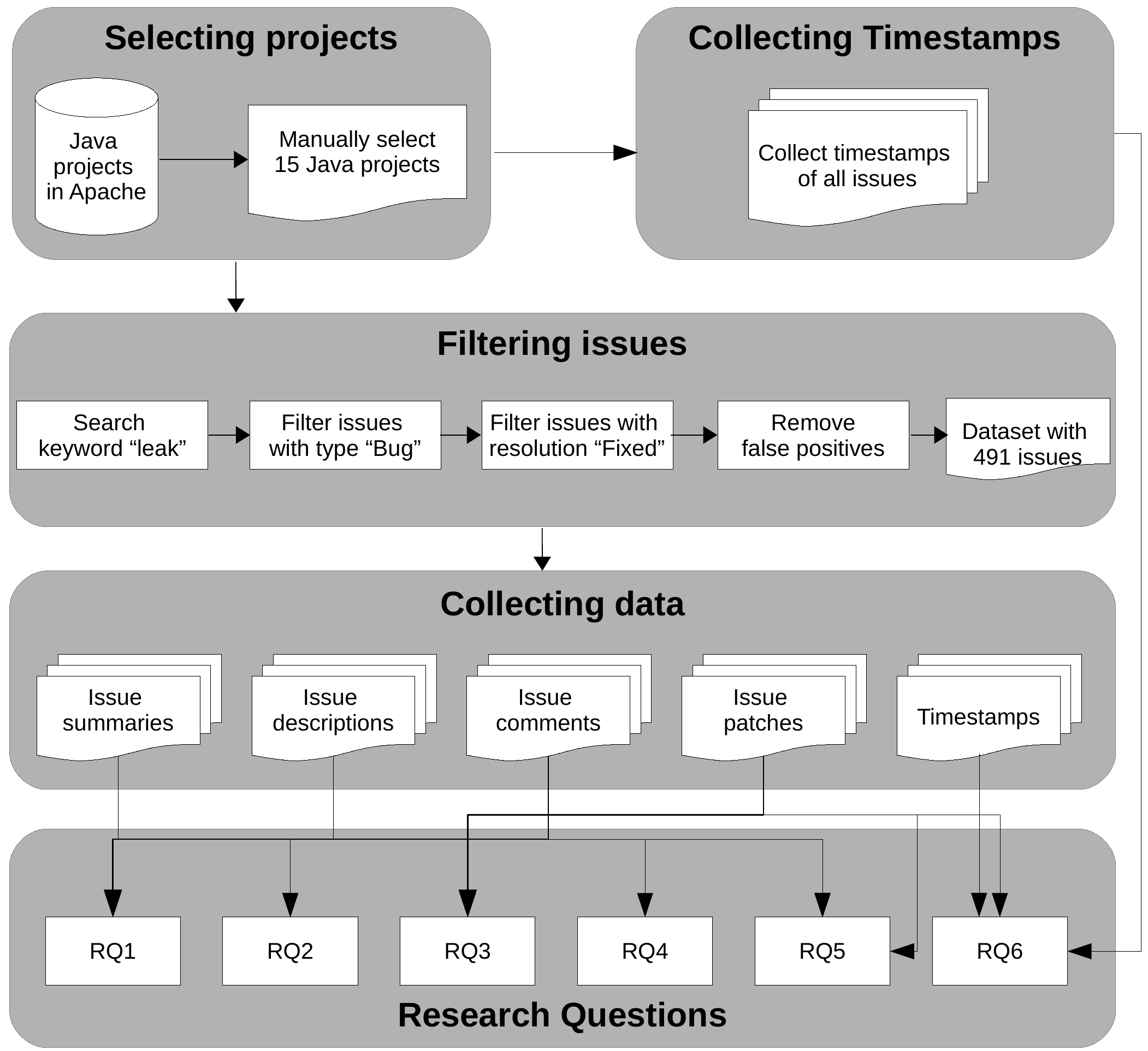}}	
\caption{Overview of the empirical study design.}
\label{fig:overview}
\end{figure}

In this section, we describe the design of our empirical study.~\Cref{fig:overview} gives an overview of our methodology. In the remainder of this section, we illustrate the research questions, studied applications, and data collection process.

\subsection{Studied Projects}
\begin{table*}[t]
\caption{Overview of studied projects. The Java LOC for each project is obtained from Open Hub.
}
\begin{tabular}{llrrr}
\hline
\textbf{Project}&\textbf{Category}&\textbf{First Commit}&\textbf{\#Committers}
&\textbf{\#kLOC}\\ 
\hline
AMQ			&Distributed messaging	&2004&58&1158\\
CASSANDRA	&Distributed database	&2009&45&313\\
CXF			&Web service 			&2007&38&674\\
DERBY		&Relational database	&2004&44&689\\
HADOOP		&Distributed computing	&2006&163&1260\\
HBASE		&Distributed database	&2007&57&1115\\
HIVE		&Data warehouse			&2009&63&1074\\
HTTPCOMP. 	&Network client/server	&2004&18&115\\
LUCENE		&Search framework		&2004&67&557\\
SOLR		&Search framework		&2008&67&416\\
Realm Java  &Mobile database        &2012&14&116\\
Spring Boot &Application framework  &2012&180&311\\
Logstash    &Data Processing        &2009&43&74.6\\
RxJava      &Reactive programming   &2013&65&279\\
Selenium    &Browser driver         &2006&115&703\\
\hline
\end{tabular}
\label{table:projectCategories}
\end{table*}

We perform a study on 15 open-source Java projects hosted in two major repositories, Apache and GitHub. We investigate the leak-related issues from a wide variety of software categories to ensure the diversity of the studied projects. \Cref{table:projectCategories} lists the studied projects. 

We study these projects for two reasons. First, they are large-scale and open-source projects with a mature codebase with years of development. We believe that by using such a well-established and well-developed applications, we can get results representative for mature Java projects. Column \#kLOC in Table~\ref{table:projectCategories} shows the size of the Java source code of the studied projects ranging between 74 to over 1200 kLOC. For the Github projects, the total number of stars is more than 100k.

Second, their issues are reported and tracked in a bug tracking system. Similar to other bug trackers (e.g., Bugzilla), reports in JIRA or GitHub bug tracker are well-described and provide sufficient information to answer the research questions investigated in this study. 

\subsection{Research Questions}
The following research questions guide our study:

\noindent \textbf{RQ1. \RQI}
In~\Cref{RQ100}, we analyze the dominant leak types in each project. We use this analysis in the next research questions to distinguish the properties of different leak types.

\noindent \textbf{RQ2. \RQII}
Understanding different detection types can help leak detection approaches to improve detection accuracy. In~\Cref{sec:detection}, we investigate how developers or users report the leak-inducing defects and how the leaks manifest at runtime. We analyze different detection and manifestation types and study their relation to the leak types.

\noindent \textbf{RQ3. \RQIII}
Bug localization is the first step in bug diagnosis. The extent of the bug can highly affect the number of files that need to be fixed to repair the bug. In this question, we analyze the locality of leak-inducing defects~(\Cref{RQ300}).

\noindent \textbf{RQ4. \RQIV}
\Cref{RQ400} describes the common root causes of leak defects. We investigate the prevalence of each root cause and their relation to the leak types.

\noindent \textbf{RQ5. \RQV}
In~\Cref{RQ500}, we identify the repair actions applied by the developers to repair the leak-related defects and investigate the frequency of each considering different leak types. We also search to find recurring code transformations in the repair patches. We identify 13 common repair patterns from our dataset. In this question, we investigate whether the automated program repair techniques (i.e., the process of providing the repair patches automatically) such as template-driven patch generation are applicable for fixing the leak-related defects. 

\noindent \textbf{RQ6. \RQVI}
In~\Cref{RQ600}, we measure the code churn, change entropy, and diagnosis time to assess the complexity of the changes needed to repair the leak-inducing defects. This analysis provides insights about the difficulty of repairing the leak-related defects and shows which type of leaks can be repaired with less effort in terms of time and amount of code changes.

\subsection{Data Extraction}
\label{sub:DatasetAllIssues}
For the Apache projects, we collected the leak-related issues from the bug tracker reported until June 2016. For GitHub projects, we collected the issues reported until January 2019.  

To build a suitable dataset for our study, we apply a four-step filtering methodology: (1) keyword search, (2) issue type filtering, (3) resolution filtering, and (4) manual investigation. This four-step filtering method yields a dataset with 491 leak-related issues, each representing a unique leak bug report (i.e., none are duplicates of another). We make the dataset available online\footnote{https://github.com/heiqs/leak\_study}.

\subRQ{Keyword search.} We use a simple heuristic and select issues that contain the keyword ``leak'' in the issue title or issue description. The keyword search is a well-known method used by previous empirical studies \citep{EmprunderstandingPerfBugs,EmprBugFix,EmprPerfBugMSR} to filter the issues of interest from the others. It is worth mentioning that we investigated other leak-related keywords (unreleased, out-of-memory, OOM, closed, etc.). However, these keywords yield a dataset with a high number of false positives. For example, the keyword “unreleased” is used in the title of the issue report CXF-7776\footnote{https://issues.apache.org/jira/browse/CXF-7776}: “Download page should not link to unreleased code”. This is obvious that this issue has no relation to this study. Pruning of such issues is time-consuming and requires a huge amount of manual effort. On the other hand, it is possible that we skip some leak-related issues due to our simple keyword search. For example, YARN-5257\footnote{https://issues.apache.org/jira/browse/YARN-5257} refers to some unreleased resources which are fixed. Although this is a leak-related issue, the term leak is not mentioned in the issue title or description.

Despite the simplicity of keyword search, this heuristic proved to be highly precise due to the high quality of issue reports and related data in the studied projects. \citet{Wu_Relink} highlight that even simple heuristics can yield the same precision and recall as more sophisticated search techniques when applied to a well-maintained bug tracker. Using the keyword search, we identify 1255 leak-related issues. Column ``\#Issues'' in~\Cref{table:issues} shows the number of filtered issues for each project. 

\subRQ{Issue type filtering.} Each issue in the bug tracker can be classified as ``Bug'', ``Task'', ``Test'', and so on. As we are only interested in leak-related bugs, we first filter issues with type ``Bug''. Among the 1255 issues filtered by keyword search, there are 838 issues labeled as a bug (column ``\#Bugs'' in~\Cref{table:issues}).

\subRQ{Issue resolution filtering.} To analyze how developers repair a leak defect we need to restrict our analysis to fixed bugs. For this, we filter issues with the resolution label ``Fixed'' for Apache projects and ``Closed'' for GitHub projects. This reduces the dataset to 591 issues (column ``\#Fixed'' in~\Cref{table:issues}).

\subRQ{Manual investigation.} In the final step, we remove the false positives from our dataset. We manually filter out the following issues:
\begin{itemize}[leftmargin=*]
\item Non-leak-related bugs retrieved by our keyword search heuristic. For instance, in issue CXF-3390\footnote{https://issues.apache.org/jira/browse/CXF-3390}, the term \textit{leak} is used in ``information leak'' which is not related to this study. 
\item Wrongly reported leaks. These issues should be tagged as ``Invalid'', but are closed in the bug tracker without correct labeling. 
\end{itemize}
\begin{table*}[t]
\caption{Studied projects with statistics on number of issues (explained in Section~\ref{sub:DatasetAllIssues}). Columns ``\#MLeak'', ``\#RLeak'', and ``Total'' show the numbers of memory and resource leak issues per application, and their totals, respectively. 
}
\begin{tabular}{lrrr|rrr}
\hline
\textbf{Project}&\textbf{\#Issues}&\textbf{\#Bugs}&\textbf{\#Fixed} &\textbf{\#MLeak}&\textbf{\#RLeak} &\textbf{Total} \\ 
\hline
AMQ			&123	&116 	&88		&54		&26		&80\\
CASSANDRA	&77		&65		&45		&19		&16     &35\\
CXF			&62		&61		&44		&29		&8		&37\\
DERBY		&50		&36		&23		&12		&4		&16\\
HADOOP		&236	&201	&132	&43		&76		&119\\
HBASE		&92		&65		&44		&11		&29		&40\\
HIVE		&78		&69		&47		&19		&25		&44\\
HTTPCOMP. 	&31		&28		&24		&8		&12		&20\\
LUCENE		&77		&65		&42		&13		&21		&34\\
SOLR		&74		&60		&33		&11		&16		&27\\
Realm Java  &76     &15     &15     &4      &2      &6\\      
Spring Boot &94     &17     &16     &2      &10     &12\\
Logstash    &67     &25     &23     &8      &4      &12\\
RxJava      &100    &14     &14     &5      &3      &8\\
Selenium    &18     &1      &1      &0      &1      &1\\
\hline
\textbf{Total}	&\textbf{1255}	&\textbf{838}	&\textbf{591}	&\textbf{238} 	&\textbf{253}	&\textbf{491}\\
\hline
\end{tabular}
\label{table:issues}
\end{table*}

\subsection{Tagging Leak-Related Defects}
\label{labeling}

To analyze the properties of the leak-related defects, we need to classify the issues for each dimension of interest (i.e., leak type, detection type, detection method, defect type, and repair type). However, we only have qualitative information such as issue description, developers discussions, and repair patches. There is no label provided in the bug tracker for classification of the attributes that we are interested in reported leaks. To derive properties for the bugs in our dataset, we need to quantify the qualitative information. For this purpose, we perform an iterative process similar to \textit{Open Coding}~\citep{Seaman99-coding,Seaman2008-coding}. In our study, the input of the coding process for each issue is issue summary, issue description, developers discussions, and repair patches. The first author of the paper (a Ph.D. student), classified a sample set of the issues to determine the possible categories for each dimension. After identifying the initial types for each category, the second and the third authors (a Ph.D. student and an undergraduate student) join the first author to discuss the categories and label the remaining issues. We held many meetings, spent many hours, and performed multiple iterations to achieve cohesive labeling. 

The tagging process is iterative. Each time a new type is identified, the coders (first three authors) verify it in a decision-making meeting. If a majority of the coders agree on the new type, they go through all the previously tagged issues and check if the issues should be tagged with the new type. This also minimizes the threat of human error during the labeling process. To further reduce the error probability and in case of difficulty in classifying of the issues, all the coders check and discuss the complex issues to find the appropriate categories. The conflicts were resolved by discussing and coming to an agreement.

To validate the manual labeling process, we apply the following procedure. The first and second author perform a classification of a statistically representative sample of the dataset with a confidence level of 95\% and a confidence interval of 10\%. This results in a sample set of 80 out of 491 issues. We calculate the inter-rater agreement with Cohen's kappa metric~\citep{kappaCohen, KappaLinguistics}.~\Cref{tab: kappa} shows the result of our analysis. The lowest Cohen's kappa value is for the repair type, although it shows a moderate agreement between the two coders. The reason for disagreements is that the categories in this attribute are not mutually exclusive. Therefore, there is a probability that each coder has a different interpretation of the same issue. After rating, the two raters discussed their disagreements to reach consensus.

\begin{table}[t]
\centering
\caption{Cohen's kappa measurement.}
\begin{tabular}{lr}
\toprule
\textbf{Dimension} & \textbf{Cohen's Kappa} \\
\midrule
Leak Type (RQ1) & 0.86 \\
Detection Type (RQ2) & 0.83 \\
Detection Method (RQ3) & 0.70 \\
Defect Type (RQ4)& 0.69 \\
Repair Type (RQ5)& 0.57 \\
\bottomrule
\end{tabular}
\label{tab: kappa}
\end{table}

\subsection{Uniqueness of categories}
\label{sec:uniqueness}
During tagging task, we encounter the issues with the possibility of assigning them to multiple categories. For example, in Hadoop-6833\footnote{https://issues.apache.org/jira/browse/HADOOP-6833}, a leak is reported due to the forgotten call to the~\textit{remove} method of a collection. The developers repaired the bug by adding the remove call in the exception path:

\begin{lstlisting}
--- src/java/org/apache/hadoop/ipc/Client.java
+++ src/java/org/apache/hadoop/ipc/Client.java
@@ -697,6 +697,7 @@ public class Client {
         } else if (state == Status.ERROR.state) {
           call.setException(new RemoteException(WritableUtils.readString(in),
                                                 WritableUtils.readString(in)));
+          calls.remove(id);
         } else if (state == Status.FATAL.state) {
           // Close the connection
           markClosed(new RemoteException(WritableUtils.readString(in),
\end{lstlisting}

One could label this issue as~\textit{collection mismanagement}. However, if the exception is thrown no leak is triggered. Therefore, we use the underlying cause as the main root-cause category (here \textit{bad exception handling}). For the repair action, we assign a bug to the category used by the developer to repair the defect. In this example, we label the repair action as~\textit{remove element}.


\section{Empirical Study Results}
\label{results}
\label{sec:results}
In this section, we answer the research questions. For each research question, we describe the motivation behind the question, the approach used in answering the research question, and the findings derived from the analysis.

\subsection{RQ1: \RQI}
\label{leakTypes}
\label{RQ100}

\subRQ{Approach.} For most of the studied issues, the reporters or developers explicitly mentioned the leak type. For such cases, we assign the leak type as reported. In case of no explicit mention of the leak type, we manually analyze the title, description, and developers discussions to assign the correct leak type.

\subRQ{Taxonomy of leak types.} Our analysis yields a taxonomy of leak types with the following four categories:
\begin{figure}[t]
\centering
\scalebox{0.34}{
\includegraphics{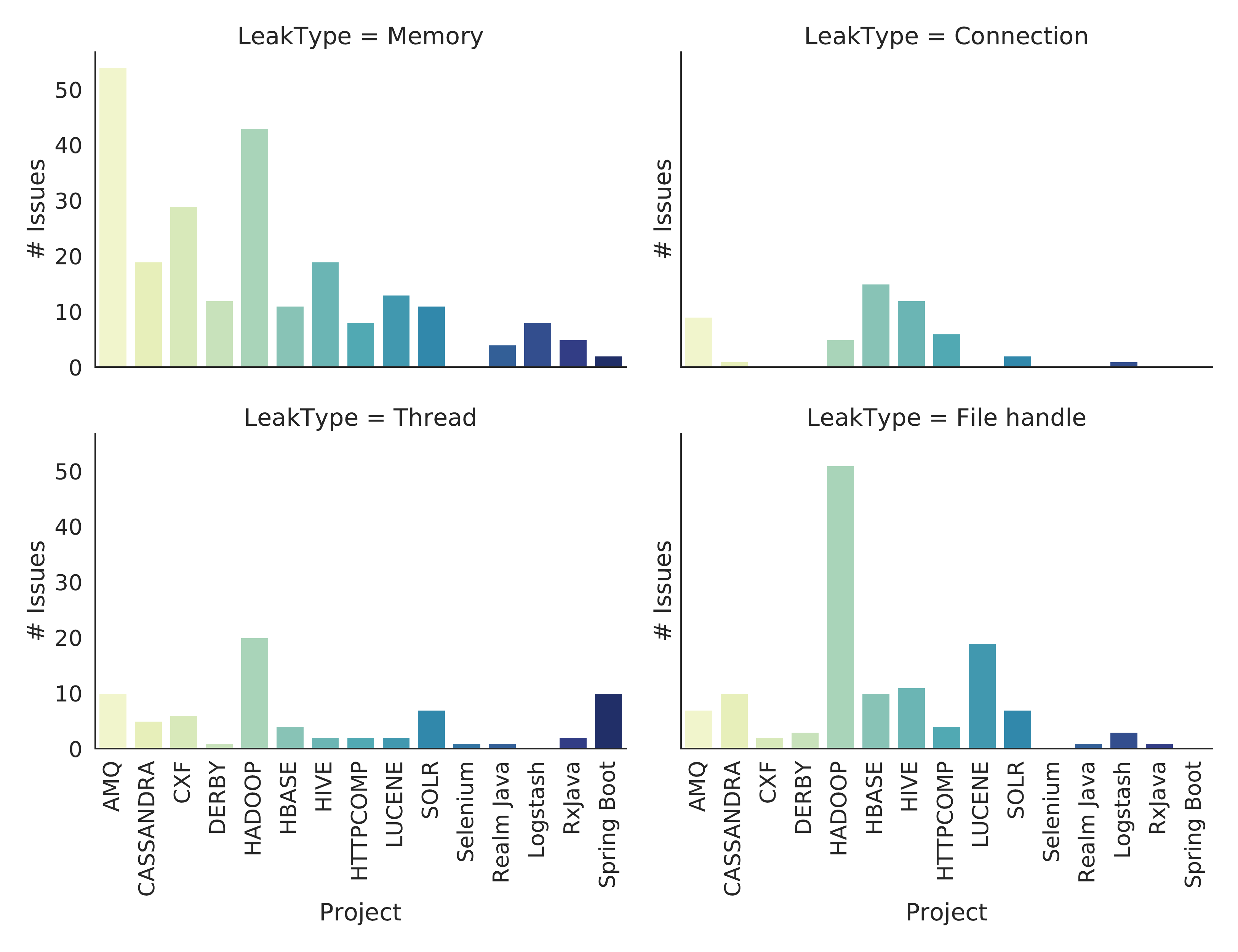}}	
\caption{Frequency of the leak types per project.}
\label{fig:leak-type}
\end{figure}

\category{Memory} We group in this category all issues reported due to unreleased references to Java objects, such as mismanagement of collections or circular references. 

\category{File handle} We group in this category leaks related to file descriptors. These issues are related to the mismanagement of Java file handlers such as I/O streams.

\category{Connection} We group in this category leaks triggered by non-closed network or database connections. 

\category{Thread} We group in this category leaks caused by unclosed, yet unused threads. A thread leak occurs when a no-longer-needed thread is unnecessarily kept alive. This thread then leaks its internal resources, which cannot be released by the JVM. 

\subRQ{Results.} ~\Cref{fig:leak-type} shows the distribution of the leak types for each project. We use this data to find the dominant leak types in the projects and in the project categories. 

\textbf{Finding 1.} The three leak types corresponding to the resource leaks (i.e., file handle, connection, and thread) is the most common leak types in six out of the ten projects. Resource leaks (with \resource issues) are slightly more reported than memory leaks (with \memory issues).

\textbf{Finding 2.} Each project shows a distinct distribution of the leak types. LUCENE and HADOOP have a higher frequency of the {\em file handle} leak type with this leak type corresponding to 55.9\% and 42.9\% of the issues, respectively. Projects AMQ (67.5\%), CASSANDRA (54.3\%), CXF (78.4\%), and DERBY (75.5\%) contain more memory leak issues. Connection leaks are more frequently reported in HBASE (37.5\%), HTTPCOMP (30\%), and Hive (27.3\%). 10 out of 12 issues in Spring Boot are of type thread leak. This analysis shows the diversity of the leak types in the studied projects. Even projects within the same category show different distributions of the leak types. 

\begin{tcolorbox}
\noindent \textbf{Summary.} Resource leaks (253 out of 491 issues) are slightly more often reported than memory leaks (238 issues). Leak type distribution is different across the projects. 
\end{tcolorbox}

\subsection{RQ2: \RQII}
\label{detection} \label{RQ200} \label{sec:detection}
\subRQ{Motivation.} Each issue report provides information about leak symptoms, environmental setup, and methods used to detect a leak. Understanding how leaks are detected can provide valuable insights on leak diagnosis. It also shows in which direction the researchers and tool builders should help programmers in leak detection. In this question, we want to find whether the leaks are detected during runtime and whether the static analysis is used for leak detection.

\subRQ{Approach.}
To find detection type for each issue, we use three data sources: issue title, issue description, and developers discussions. Using this data, we analyze the frequency of the detection types, distribution of detection methods, and their relation to different leak types.

\subRQ{Taxonomy of leak detection.}
Leak-inducing defects can be discovered with and without runtime failures or performance degradation. They can be detected via manual analysis of the source code, an unexpected runtime failure (in particular, an out-of-memory error), or abnormal usage of resources. We classify detection types into two major categories: source code-based detection and runtime detection. In the following, we explain these two detection types in more detail.

\category{Source code-based detection} In this category, we classify issues such that the leak detection is performed before execution of the program and there is no reported runtime information in the issue report, nor reports on leak-related failures. We observe that issue reporters describe these issues with phrases such as ``can potentially cause a leak'' or ``can lead to a leak''. The main techniques to detect leaks prior to the runtime are \textit{manual code inspection} and \textit{static analysis tools}.

Manual inspection of the source code (or code review) is a process in which developers inspect a set of program elements (e.g., methods, classes) in order to improve the quality of software~\citep{BestPractices, McConnellCodeComplete, SommervilleSoftEng}. It is one of the most common static detection methods used by developers in practice. This detection type requires the knowledge of how a leak can be introduced as well as an understanding of the application's behavior. For instance, in AMQ-5745\footnote{https://issues.apache.org/jira/browse/AMQ-5745}, manual inspection revealed cases where bad exception handling could yield resource leaks on the AMQ codebase. 

Static analysis tools can be used to identify potential leak defects during the development process. 
There are many free and proprietary static analyzers which are able to detect specific leak types (e.g., FindBugs, Infer). 

\begin{figure}[t]
\centering
\scalebox{0.4}{
\includegraphics{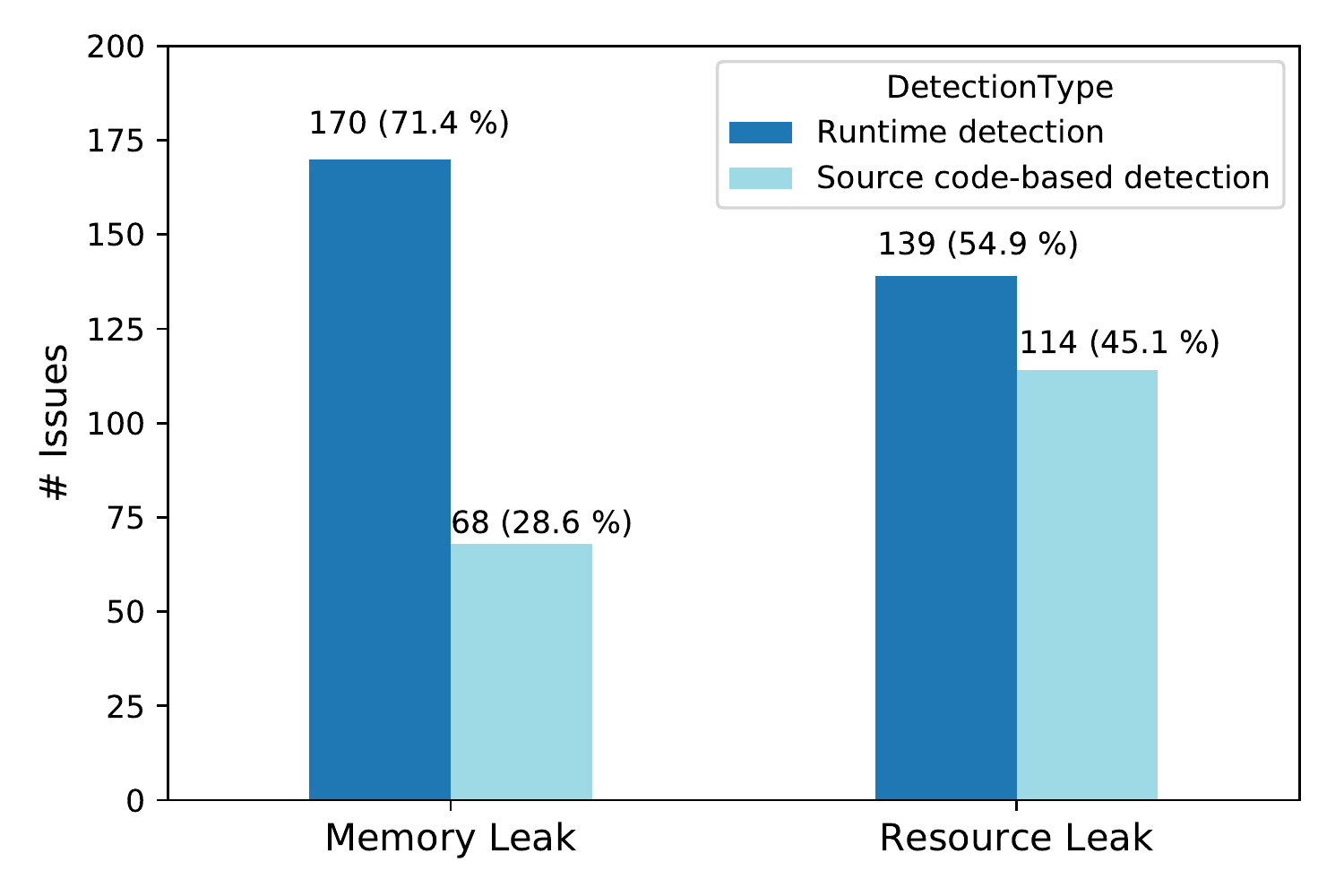}}	
\caption{Frequency of the detection types per leak type.}
\label{fig:detection-type}
\end{figure}

\category{Runtime analysis} Some leak-related failures are observed and reported when a user/developer encounters a performance degradation in a production environment, an out-of-memory error is raised, or a test is failed. Issue reporters often use phrases such as ``consistently observing memory growth'' or ``meet memory leak in a production environment''. In these issues, the bug reporter often provides additional material such as heap profile, memory dump, a log file, or a stack trace. This supplementary data can help developers on localizing the root cause of the leak defect more efficiently. Leaks usually manifest in the runtime with a symptom. From our observation, we identify three symptoms reported in the issue reports: failing tests, out-of-memory errors, and warning messages.

The output of a failing test case may expose a leak. The test can be a system test, a unit test or any other application-provided test. For example, in LUCENE-3251\footnote{https://issues.apache.org/jira/browse/LUCENE-3251}, a failing unit test exposed a file-handle leak. The user provided the stack trace of the failing test in the issue report:

\begin{small}
\begin{lstlisting}
Testsuite: org.apache.lucene.index.TestAddIndexes
Testcase: testAddIndexesWithRollback(org.apache.lucene.index.TestAddIndexes):
  Caused an ERROR
    MockDirectoryWrapper: cannot close: there are still open files: {_co.cfs=1}
    java.lang.RuntimeException: MockDirectoryWrapper: cannot close: there are still open files: {_co.cfs=1}
      at org.apache.lucene.store.MockDirectoryWrapper.close(MockDirectoryWrapper.java:483)
\end{lstlisting}
\end{small}

In some cases, the growth of memory usage leads to an out-of-memory (OOM) error during runtime. This is a severe symptom as the underlying system often crashes when an OOM error occurs. For example, DERBY-5730\footnote{https://issues.apache.org/jira/browse/DERBY-5730} reported a severe memory leak which might lead to a system crash due to an out-of-memory error. In this issue, the reporter mentioned that after removing the suspicious call, the test program was successfully executed with a much lower heap size. We should mention that we only consider those out-of-memory errors triggered via a resource or memory leak and not because of misconfigured heap size (which is a configuration error). The out-of-memory error due to a leak occurs at some point regardless of the heap size, while the OOM error due to a misconfiguration will not occur with correct configuration of the heap value. Logstash\#5179\footnote{https://github.com/elastic/logstash/issues/5179} is an example of an OOM error due to a misconfiguration of the heap size for running a specific task which is fixed by setting the correct value for the heap size.

Developers also implement algorithms for detection of specific leak defects. They usually warn the user about the potential presence of a leak with a message during program's execution. For example, in CXF-5707\footnote{https://issues.apache.org/jira/browse/CXF-5707}, a message warned the user for a potential leak during the performance test of the \textit{netty-http-server} module:

\vspace{10pt}
\begin{tabular}{|p{{0.8}\textwidth}}
''\textit{SEVERE: LEAK: ByteBuf.release() was not called before it's garbage-collected. Enable advanced leak reporting to find out where the leak occurred.}''
\end{tabular}
\vspace{10pt}

\begin{table}[t]
    \centering
    \caption{Distribution of detection methods for memory and resource leaks.}
    \label{Tab:detection-relation}
    \begin{tabular}{llrr}
    \toprule
                     & &  \textbf{Memory Leaks}&  \textbf{Resource Leaks}\\
    \textbf{Detection Type} & \textbf{Detection Method} &         &           \\
    \midrule
    Source code-based & Manual code inspection &    \percentwithmaxbar{68}{238} &     \percentwithmaxbar{113}{253} \\
     detection  & Static analyzer &     \percentwithmaxbar{0}{238} &       \percentwithmaxbar{1}{253} \\
    \midrule
                & Failed test   &    \percentwithmaxbar{17}{238} &      \percentwithmaxbar{40}{253} \\
    Runtime detection   & Out-of-memory error &    \percentwithmaxbar{43}{238} &      \percentwithmaxbar{12}{253} \\
       & Warning message &     \percentwithmaxbar{8}{238} &       \percentwithmaxbar{11}{253} \\
                & Runtime (exclude above) &    \percentwithmaxbar{102}{238} &      \percentwithmaxbar{76}{253} \\
    \bottomrule
    \end{tabular}
\end{table}


\subRQ{Results.}
\Cref{fig:detection-type} shows the distribution of detection types in relation to the leak types. \Cref{Tab:detection-relation} illustrates the relationship between detection types, detection methods, and leak types.

\textbf{Finding 1.} More resource leaks (114 issues) are detected via source code-based detection than memory leaks (68 issues). Runtime detection is the dominant detection type for detecting memory leaks with 170 out of 238 issues (71.4\% of the issues). The reason why more resource leaks are detected with source code-based detection techniques comes from the difference in memory and resource management. In Java, a programmer should explicitly dispose of the resources after usage. Due to explicit management, potential resource leaks can more often be captured through the code review or using static analyzers. Contrary to this, the JAVA Virtual Machine (JVM) manages the heap space and releases the unused objects when they become unreachable. Detecting unused references with code inspection is a hard task, as the programmer needs to have a profound understanding of the program's workflow.

\textbf{Finding 2.} 309 (about 63\%) issues are detected or manifest during the runtime. In these issues, users often use a third-party memory analyzer (e.g., \emph{jmap}, \emph{MAT}\footnote{https://www.eclipse.org/mat/}, \emph{yourkit}\footnote{https://www.yourkit.com/}), or OS-specific commands (e.g., \emph{lsof}) to verify the presence of the leaks. The information collected from these tools and commands can considerably help the developers to reproduce and diagnose the leak defects.

\textbf{Finding 3.} Users detected leaks in 19 issues (3.9\%) via warning messages. From our dataset, we observe that in three applications, developers implemented leak detection mechanisms. This result shows that it is a good practice for developers to provide integrated leak detection mechanisms to accelerate the diagnosis of leak-related defects. 

\textbf{Finding 4.} Out-of-memory errors are observed more than three times in memory leak-related issues. OOM error is one of the most severe leak symptoms and should be particularly prevented in a production environment.

\textbf{Finding 5.} In 57 (11.6\%) issues, users detected the leaks via a test case. We also observe that for some issues, developers added a test case to the repair patches for future leak detection. This result shows the possibility of software tests as a lightweight tool for leak detection. Previous work~\citep{perfblower, SCAM2015} confirm the effectiveness of software tests for leak detection. The utility of a test case is twofold. First, it can be used as an oracle for automated leak detection and bug isolation. Second, it can be an oracle for automated leak repair techniques as they need test cases to verify the correctness of their proposed fix patches. As leaks are highly environment - and input - sensitive, the automated test input generation should provide inputs which can trigger the leaks in different execution paths.

\textbf{Finding 6.} Only in one issue (CASSANDRA-7709\footnote{https://issues.apache.org/jira/browse/CASSANDRA-7709}), the leak is detected and reported by a static analyzer. As we only consider the reported issues, we cannot generalize that the static analysis tools are not used. It is possible that static analyzers have been employed earlier in the development process, and all leaks detected were fixed. Still, our finding highlights potentials to improve the existing static analyzers further as there is still room for improvement. It is important to know why these tools are not used for other reported issues with similar characteristics to the detected issue (we further analyze this in \Cref{sec:others}). One reason might be that there are still obstacles in the extensive use of such debugging tools. Such obstacles can be high false positives, complex usage procedure, or lack of knowledge about these tools. Researchers or tool builders should improve current debugging tools to detect not-yet covered bugs, simplify the tool usage, and spread them widely in the community.

\begin{tcolorbox}
\noindent\textbf{Summary.} Source code-based detection is more common in resource leak detection (45.1\%). Runtime detection is the dominant detection type for memory leaks (71.4\%). Out-of-memory errors are observed about three times more frequently in conjunction with the memory leaks than with the resource leaks.
\end{tcolorbox}

\subsection{RQ3: \RQIII}
\label{sub:localization}
\label{RQ300}
\subRQ{Motivation.} Fault localization is the first step of bug diagnosis. The locality of a fault can be defined in different granularity such as statement, method, and file. In the case of leak-related defects, they can affect a region (e.g., multiple modules, classes, etc.) in the codebase of an application~\citep{Leakbot2003}. Accurate defect localization is vital in providing the correct repair patch. Otherwise, the patch will not fix the bug completely and even introduces a new bug~\citep{fixBecomeBug}. In this research question, we investigate the locality of leak-inducing defects. In particular, we want to find out: (1) how many source code files are changed to repair a defect, and (2) which types of files are changed in each repair patch.

%

\subRQ{Approach.} To assess the locality of leak defects, we analyze the distribution of modified source code files. For each issue, we collect the files changed in the repair patches. We also investigate the file type of modified source files by collecting the file extensions.
We ignore test files in the repair patches if the tests added or modified for future leak detection and not for repairing purposes. 

\begin{figure}[t]
\centering
\scalebox{0.55}{
\includegraphics{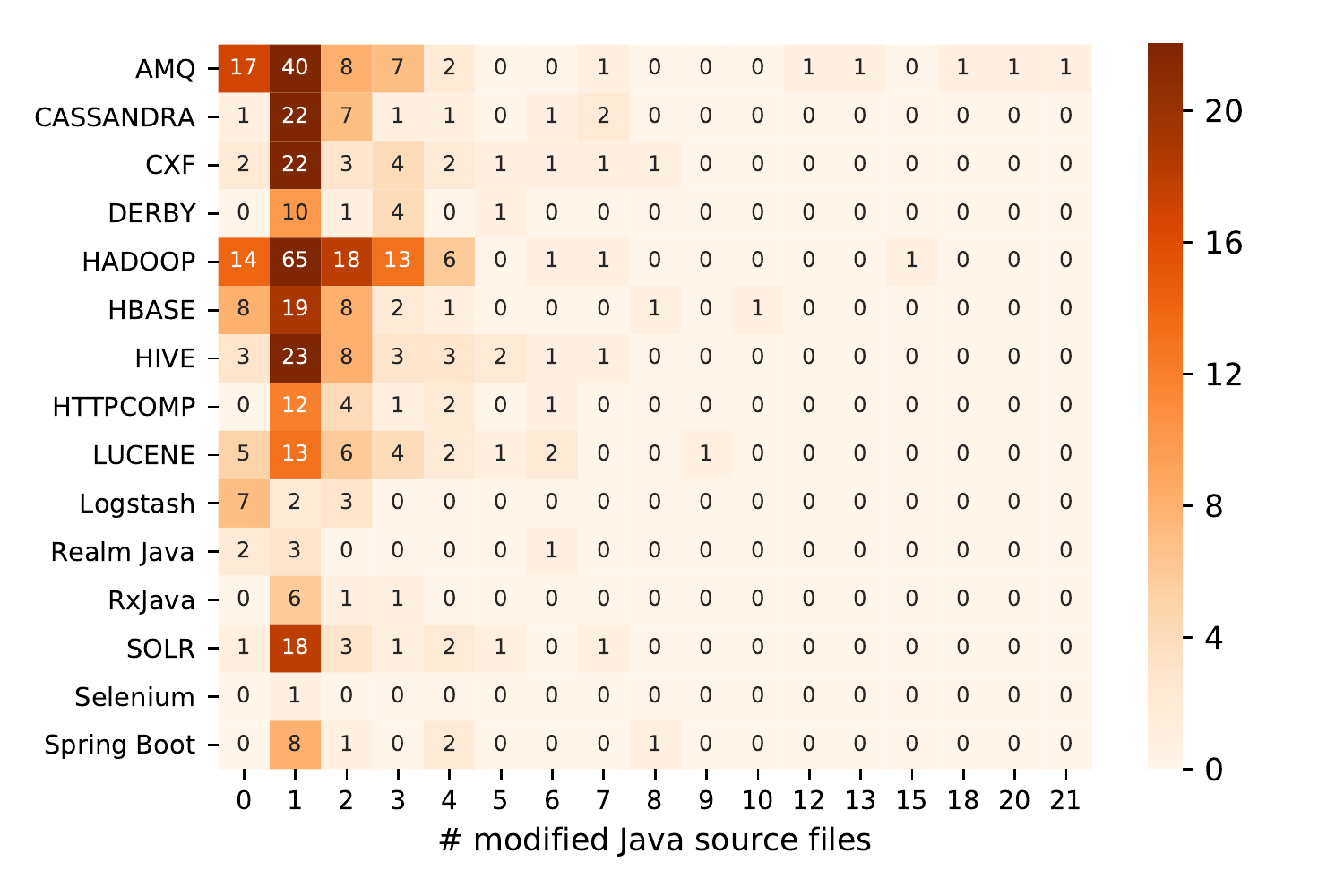}}
\caption{Heatmap of the number of modified Java source code files per project.}
\label{fig:heatmapFileCount}
\end{figure}

\subRQ{Results.}~\Cref{fig:heatmapFileCount} shows the heatmap representation of the amount of modified Java source files for each project.

\textbf{Finding 1.}
In 57\% of the issues, developers changed only one source code file to repair the defect. In about 81\% of the issues, three source code files are modified at most. This result implies the high locality of leak-inducing defects considering file-level granularity. 

\textbf{Finding 2.}
In 15 issues, developers repaired the defect via adding or deleting at least one Java source code file. It is interesting to know the reasoning behind such changes. However, the information for such decisions is not always provided in the issue reports and reasoning require deep knowledge about the design and architecture of the project. Our further analysis shows that the decision of adding or deleting a class is simply a design decision and it is very particular to the issue being fixed. Here, we provide three examples of such cases. In CASSANDRA-552\footnote{https://jira.apache.org/jira/browse/CASSANDRA-552} developer created a new interface (in a new file) which makes an iterator object to be closable. In HADOOP-639\footnote{https://jira.apache.org/jira/browse/HADOOP-639} developers redesigned the code with unifying two existing functionalities. In LUCENE-1383\footnote{https://jira.apache.org/jira/browse/LUCENE-1383} developers implemented a closable Java \texttt{ThreadLocal} as a wrapper for Java \texttt{ThreadLocal} which wraps the values inside a \texttt{WeakReference} with keeping a strong reference to the value. In this way, the garbage collector does not reclaim the value until the close method is called. In general, we could not generalize any specific rule about adding or deleting files in the repair patches of the studied issues.

Albeit occurring only in about 3\% of the studied issues, such cases require more sophisticated repair strategies. Most of the current automated program repair approaches~\citep{genprog,templateFix,HDRepair} can only provide simple repair patches with only one line. Hence, it is still not feasible for existing automated program repair techniques to provide complex repair patches such as adding a class.

\textbf{Finding 3.}
In 17 issues, no Java source code file is changed. In eight issues, source code files written in C are modified. In three cases, developers modified the XML files to use a non-leaky third-party library as a dependency for that project. 6 issues are repaired by changing source code files written in Scala and Ruby. The reason for changing different file types is that in some of the studied projects, specific modules are implemented in different programming languages than Java. For example, bzip2 (a compression method) implementation in HADOOP is written in C. 

Test cases might also contain a leak in their code. For example, YARN-2662 reports an issue where the tests do not close a file after reading from it. We observe 15 issues in our dataset that the repair patches contain only changes in the test files.

A bug report is labeled as duplicated if it has already been reported in the bug tracker. However, it can be the case that a reporter reports a bug and this bug is already repaired in one of the previous releases of the software or the root cause of the leak is located in a third-party library. In such cases, developers close the issue as fixed with referring to the software release containing the bug fix or the non-leaky third-party library. In our study, we find 29 issues of such cases, i.e., the issues are closed without including a repair patch. It is worth mentioning that these issues cannot be considered as duplicated because they are not previously been reported in the bug tracker (i.e., there is no link to another issue in the bug tracker).
 
Although only a few defects are repaired by modifying files written in other programming languages, developers require knowledge of different programming languages to repair all leak-related defects.


\begin{tcolorbox}
\noindent\textbf{Summary.} About 54\% of the leak defects are repaired by changing only one source code file. Only in 12\% of the reported leaks, more than three source files were modified. In about 6\% of the issues, files from other languages, such as C, Scala, and Ruby are modified to fix the leak-related defect.
\end{tcolorbox}

\subsection{RQ4: \RQIV} 
\label{defect}
\label{RQ400}
\label{Sec:Defect}
\subRQ{Motivation.}
In this research question, we want to find out which root causes are dominant, and whether there are significant differences in the common root causes for different leak types. 


\subRQ{Approach.} To find the root cause, we use three data sources for each issue: issue title, issue description, and developers discussions. The categories for root cause are not mutually exclusive. For issues with the possibility of assignment to multiple categories, we select the most specific category as explained in~\Cref{sec:uniqueness}. 


\subRQ{Taxonomy of defect types.}
\Cref{tab:defectTypes} lists the taxonomy of the defect types. We describe the most common root causes here.

\category{Non-closed resource} The programmer should close any system resources such as file handles, connections, and threads when they are no longer needed. Otherwise, a resource leak is likely. For example, in HBASE-12837\footnote{https://issues.apache.org/jira/browse/HBASE-12837}, \texttt{zookeeper} connections created in the constructor of \texttt{ReplicationAdmin} left unclosed.

\category{Bad exception handling} According to Java documentation~\footnote{https://docs.oracle.com/javase/tutorial/essential/exceptions/definition.html}, an exception is an event which disrupts the normal flow of the program's instructions. When an exception is thrown, any resources accessed during the normal execution of the program remain open. If a programmer does not properly handle the exceptions, a leak is prone to happen, as shown in the following quote from an issue report:

\vspace{10pt}
\begin{tabular}{|p{{0.8}\textwidth}}
\textit{``Programmer should handle the exception properly instead of swallowing it.''
}\end{tabular}
\vspace{10pt}

For instance, in LUCENE-3144\footnote{https://issues.apache.org/jira/browse/LUCENE-3144}, \texttt{FreqProxTermsWriter} leaks open file handle if an exception is thrown during \texttt{flush()}. 

\category{Collection mismanagement} The mismanagement of elements in a collection can result in memory leak. Such leak occurs when a programmer assumes that the garbage collector collects all unused objects, even if they are still referenced. Leaks due to collection mismanagement can lead to severe memory waste, in particular when the collection is used as a static member. The reason is that the static fields are never garbage-collected. Issue YARN-5353\footnote{https://issues.apache.org/jira/browse/YARN-5353} reports a severe memory leak due to keeping the tokens in the \texttt{appToken} map of the \texttt{ResourceManager} even after task completion. 

\category{Concurrency defect} A leak can be caused by a race condition preventing the disposal of a resource or releasing references to unused objects. Issue LUCENE-6499\footnote{https://issues.apache.org/jira/browse/LUCENE-6499} reports a file handle leak if files are concurrently opened and deleted.
\begin{table*}[t]
\centering
\caption{Taxonomy of root causes. 
Column ``\#Issues'' states the total number of issues per root cause.}
\begin{tabular}{lr}
\hline
\textbf{Description (Short)} 
& \textbf{\#Issues}\\
\hline
\rowcolor{LightCyan}
\databar{Non-closed resource at error-free execution (\textbf{nonClosedRes})}{149}
& \percent{149}\\
\rowcolor{LightCyan}
\databar{Object not disposed of if exception is thrown (\textbf{exception})}{98}
& \percent{98}\\
\rowcolor{LightCyan}
\databar{Dead objects referenced by a collection (\textbf{collection})}{93}
& \percent{93}\\
\rowcolor{LightCyan}
\databar{Unreleased reference at error-free execution (\textbf{unreleasedRef})}{59}
& \percent{59}\\\rowcolor{LightCyan}
\databar{A race condition defect (\textbf{concurrency}) }{18}
& \percent{18}\\\rowcolor{LightCyan}

\databar{Wrong call schedule of disposal method (\textbf{callSchedule})}{16}
& \percent{16}\\\rowcolor{LightCyan}

\databar{Over-sized cache or buffer (\textbf{cache})}{14}
& \percent{14}\\\rowcolor{LightCyan}

\databar{Incorrect API usage (\textbf{wrongAPI})}{10}
& \percent{10}\\\rowcolor{LightCyan}

\databar{Unreleased reference due to thread-local variable (\textbf{threadLocal})}{10}
& \percent{10}\\\rowcolor{LightCyan}

\databar{Classloader keeps a bi-directional reference to a class (\textbf{classloader})}{9}
& \percent{10}\\\rowcolor{LightCyan}

\databar{Leaks related to Java native interface (\textbf{jni})}{8}
& \percent{8}\\\rowcolor{LightCyan}

\databar{Leak inside a third-party library (\textbf{leakyLib})}{7}
& \percent{7}\\

\hline
\end{tabular}
\label{tab:defectTypes}
\end{table*}

\begin{figure}[t]
\centering
\scalebox{0.55}{
\includegraphics{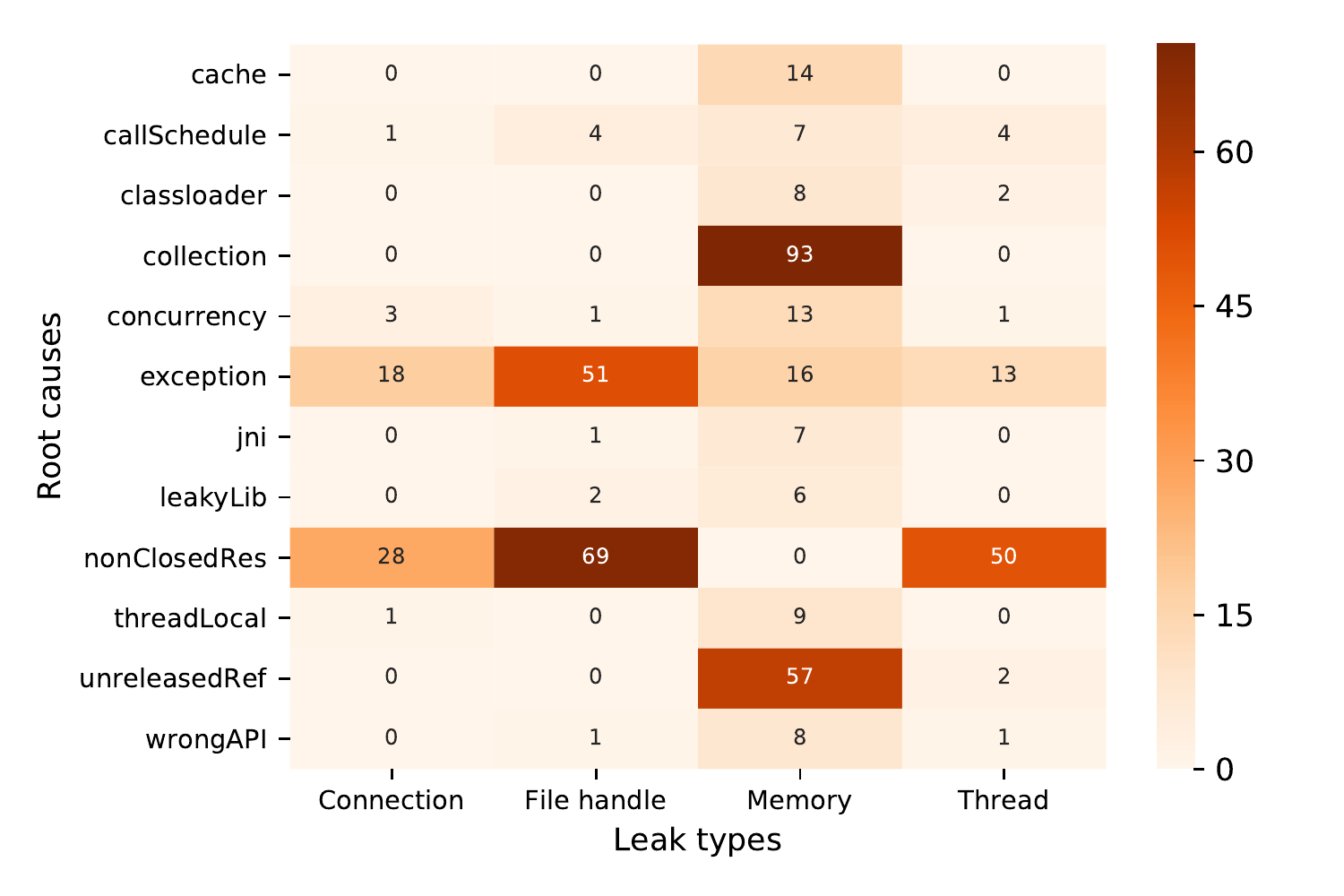}}
\caption{Heatmap of defect types and leak types.}
\label{fig:heatmapDefectLeak}
\end{figure}

\subRQ{Results.}
We investigate the frequency of the root causes across the leak types.~\Cref{tab:defectTypes} and \Cref{fig:heatmapDefectLeak} summarize the results.~\Cref{tab:defectTypes} lists the common root causes and their corresponding number of issues.~\Cref{fig:heatmapDefectLeak} visualizes the heatmap of the defect and leak types. 

\textbf{Finding 1.} The majority of the defects (about 76\% of the cases) manifest when a normal execution path is exercised. The most common root cause is also the non-closed resource in a regular (error-free) execution path (\textit{nonClosedRes}) with about 30\% of the cases. This finding is interesting. The error-free execution paths are more often executed and checked. Therefore, it should be less likely for defects to manifest in normal execution paths~\citep{weimarTemporalSpecError}. However, our analysis shows that this is not the case for leak-related defects. Our analysis shows the value of software tools and tests which check whether resources are disposed of at the end of the normal execution paths.



\textbf{Finding 2.} Bad exception handling (\textit{exception}) is the second-most frequent root cause with 20\% of the issues (93 issues). This even increases to about 32\% of the issues if we only consider resource leaks. We also observe that this root cause is about 5 times more common for resource leaks than for memory leaks. One reason for such observation is that - by definition - exception paths happen in exceptional situations, being less frequently tested than normal execution paths. Even correctly-behaved programs in normal execution path can manifest error in exceptional paths~\citep{weimarTemporalSpecError, weimarErrrorHandling}. This observation implies that the proper exception handling plays an important role in preventing leaks especially resource leaks.

\textbf{Finding 3.} Collection mismanagement (\textit{collection}) is the most common root cause for memory leaks (39\% of the cases). This finding verifies the applicability of existing automated approaches for detecting memory leaks caused by collection mismanagement (e.g.,~\citet{leakContainerXu2008}).


\begin{tcolorbox}
\noindent\textbf{Summary.} Most leaks are caused by four root causes. Collection mismanagement (39\% of the issues) and non-closed resources (58\% of the issues) are the dominant root causes for memory and resource leaks, respectively. The majority of the leaks (76\% of the cases) manifest in the regular execution paths.
\end{tcolorbox}
\subsection{RQ5: \RQV}
\label{fix}
\label{RQ500}
\label{Sec:Fix}
\subRQ{Motivation.} One approach for automated program repair is to search for common repairs from previous fix patches and provide repair candidates to fix bugs~\citep{templateFix, HDRepair, R2Fix, JSPerformance, EmpPerfStat}. Align with this direction, we investigate the repair patches to check whether there are common patterns for fixing the leak-inducing defects.

\subRQ{Approach.} For each issue in our dataset, we manually check the issue summary, the issue description, the developer discussions, and the repair patches to understand the design philosophy of a fix and find the repair action and corresponding code transformation for each defect. A repair action corresponds to an abstract description of a fix, while a code transformation is a concrete instantiation of the repair action. The same abstract fix can be implemented via different code transformations. For example, to fix a leak due to a non-closed resource (a defect type), the developer needs to dispose of the resource (a repair action). Disposing of a resource can be implemented using a \texttt{try-finally} construct (a possible code transformation) or a \texttt{try}-with-resources construct (another possible code transformation).

When analyzing the patches, we apply the following considerations. First, we are only interested in the defects within the codebase of the application. Hence, we ignore modifications of the test files in the repair patches. Second, in 29 issues, the defects are already repaired by developers in another version of the application but were not tagged as ``duplicate'' in the bug tracker. We ignore these issues for searching for common code transformation. Every label is attributed to a specific repair action whenever possible. We categorize the fix patch to a generic category only when no specific repair action would fit the repair description. 

To identify the common code transformations that may be applied for fixing multiple issues, we use the open coding methodology. First, we label each repair patch with all code transformations associated with that patch. Then, we identify those common transformations that occur repeatedly (more than once) within our dataset.

\begin{table*}[t]
\centering
\caption{Taxonomy of repair actions. 
Column ``\#Issues'' states the total number of issues per repair action.}
\begin{tabular}{lr}
\hline
\textbf{Description(Short)} 
& \textbf{\#Issues} \\ 
\hline
\rowcolor{LightCyan}
\databar{R1:
Dispose of resource in regular execution paths (\textbf{disposeReg})}{111}
& \percent{111} \\\rowcolor{LightCyan}
\databar{R2:
Dispose of objects in exceptional path (\textbf{disposeExcep})}{97}
& \percent{97} \\\rowcolor{LightCyan}
\databar{R3:
Remove the elements from a collection (\textbf{removeElm})}{104}
& \percent{104} \\\rowcolor{LightCyan}
\databar{R4:
Release the reference (\textbf{releaseRef})}{69}
& \percent{69} \\\rowcolor{LightCyan}
\databar{R5:
Shutdown thread after finishing the task (\textbf{threadDown})}{45}
& \percent{45} \\\rowcolor{LightCyan}
\databar{R6: 
Improve thread safety by avoiding race condition (\textbf{threadSafe})}{23} 
& \percent{23} \\\rowcolor{LightCyan}
\databar{R7:
Use an efficient API to improve memory usage (\textbf{correctAPI})}{12}
& \percent{12} \\\rowcolor{LightCyan}
\databar{R8:
Modify strong reference to a weak reference (\textbf{weakRef})}{9} 
& \percent{9} \\\rowcolor{LightCyan}
\databar{R9:
Use a non-leaky Library (\textbf{nonLeakyLib})}{4}
& \percent{4} \\\rowcolor{LightCyan}
\databar{R10:
Bugs not belonging to the above categories (\textbf{others})}{17}
& \percent{17} \\\rowcolor{LightCyan} \hline
\end{tabular}
\label{tab:repair-actions}
\end{table*}

\subRQ{Taxonomy of repair actions.}
~\Cref{tab:repair-actions} lists the repair actions. Note that the repair actions are mutually exclusive. For issues with the possibility of assignment to multiple categories, we select the most specific category as explained in \Cref{sec:uniqueness}. We describe the prevalent actions here.

\category{Dispose of resource on a regular path (disposeReg)} Resource leak defects introduced in \textit{regular} execution paths can be resolved via simply calling the disposal method after the resource usage.
In Java, this is achieved by calling the \texttt{close} \texttt{dispose} method. For example, in HADOOP-7090\footnote{https://issues.apache.org/jira/browse/HADOOP-7090}, the developer refers to closing the I/O streams in a \code{finally} block as a \textit{good practice}. Following is a partial patch for this issue:

\begin{lstlisting}
--- org/apache/hadoop/io/BloomMapFile.java
+++ org/apache/hadoop/io/BloomMapFile.java
@@ -186,10 +186,17 @@
     @Override
     public synchronized void close() throws IOException {
       super.close();
-      DataOutputStream out = fs.create(new Path(dir, BLOOM_FILE_NAME), true);
-      bloomFilter.write(out);
-      out.flush();
-      out.close();
+      DataOutputStream out =null;
+      try {
+        out = fs.create(new Path(dir, BLOOM_FILE_NAME), true);
+        bloomFilter.write(out);
+        out.flush();
+        out.close();
+        out = null;
+      } finally {
+        IOUtils.closeStream(out);
+      }
\end{lstlisting}

\category{Release reference} Any unused object in Java should be reclaimed by GC. If this object is still reachable by a live object, GC will not release its memory footprint. In such cases, the responsibility lies on the programmer to release the references preventing the object for being garbage collected (e.g., by nullifying the references to the unused objects). For example, HBASE-5141\footnote{https://issues.apache.org/jira/browse/HBASE-5141} reports a memory leak due to keeping references, even the corresponding task is finished. The fix patch nullifies the no-longer-needed objects. Following is the partial patch:
\begin{lstlisting}
--- org/apache/hadoop/hbase/monitoring/MonitoredRPCHandlerImpl.java	
+++ org/apache/hadoop/hbase/monitoring/MonitoredRPCHandlerImpl.java
@@ -217,6 +217,13 @@
...
+  @Override
+  public void markComplete(String status) {
+    super.markComplete(status);
+    this.params = null;
+    this.packet = null;
+  }
+
\end{lstlisting}

\category{Proper exception handling (disposeExcp)} Programmer should dispose of the objects or resources in all \textit{exceptional} execution paths. Otherwise, a leak is likely to happen when an exception is thrown. Issue AMQ-3052\footnote{https://issues.apache.org/jira/browse/AMQ-3052} reports a memory leak in \code{securityContexts}. It occurs when the \code{addConnection()} fails after a successful authentication check. The developer fixed the bug via adding a \code{try-catch} block and calling disposal method in the \code{catch} block:

\begin{lstlisting}
--- org/apache/activemq/security/SimpleAuthenticationBroker.java	
+++ org/apache/activemq/security/SimpleAuthenticationBroker.java
@@ -92,7 +92,13 @@
    ...
-   super.addConnection(context, info);
+   try {
+       super.addConnection(context, info);
+   } catch (Exception e) {
+       securityContexts.remove(s);
+       context.setSecurityContext(null);
+       throw e;
+   }
\end{lstlisting}

\category{Remove an element from a collection (removeElm)} No longer needed members of a collection should be removed by the programmer, allowing the garbage collector to reclaim the memory. A common repair action is the call of \textit{remove()} method of a collection to clear useless elements from being referenced by the collection object. For example, in issue YARN-3472\footnote{https://issues.apache.org/jira/browse/YARN-3472}, already expired and removed tokens are not removed from \texttt{allTokens} map resulting in a potential memory leak. Developer fixed the issue by adding a call to \texttt{remove} method which removed the expired token from the map.

\begin{lstlisting}
--- org/apache/hadoop/yarn/server/resourcemanager/security/DelegationTokenRenewer.java
+++ org/apache/hadoop/yarn/server/resourcemanager/security/DelegationTokenRenewer.java
@@ -577,6 +577,7 @@ private void requestNewHdfsDelegationTokenIfNeeded(
    ...
    if (t.token.getKind().equals(new Text(``HDFS_DELEGATION_TOKEN''))) {
        iter.remove();
+       allTokens.remove(t.token);
    ...
\end{lstlisting}

\category{Shutdown finished thread (threadDown)} A live thread of the application should be destroyed by the programmer when the thread task is completely finished. Adding a call to the shutdown method or adding a disposal method are the common repair actions for fixing the leaks caused by threads. HDFS-9003\footnote{https://issues.apache.org/jira/browse/HDFS-9003} reports a thread leak when a standby \texttt{NameNode} initializes the quota. Here, the thread pool is not shut down. To fix this bug, the developers added a call to the shutdown method.

\begin{lstlisting}
--- org/apache/hadoop/hdfs/server/namenode/FSImage.java
+++ org/apache/hadoop/hdfs/server/namenode/FSImage.java
@@ -880,6 +880,7 @@ static void updateCountForQuota(BlockStoragePolicySuite bsps,
         root, counts);
     p.execute(task);
     task.join();
+    p.shutdown();
\end{lstlisting}
\begin{figure}[t]
\centering
\scalebox{0.55}{
\includegraphics{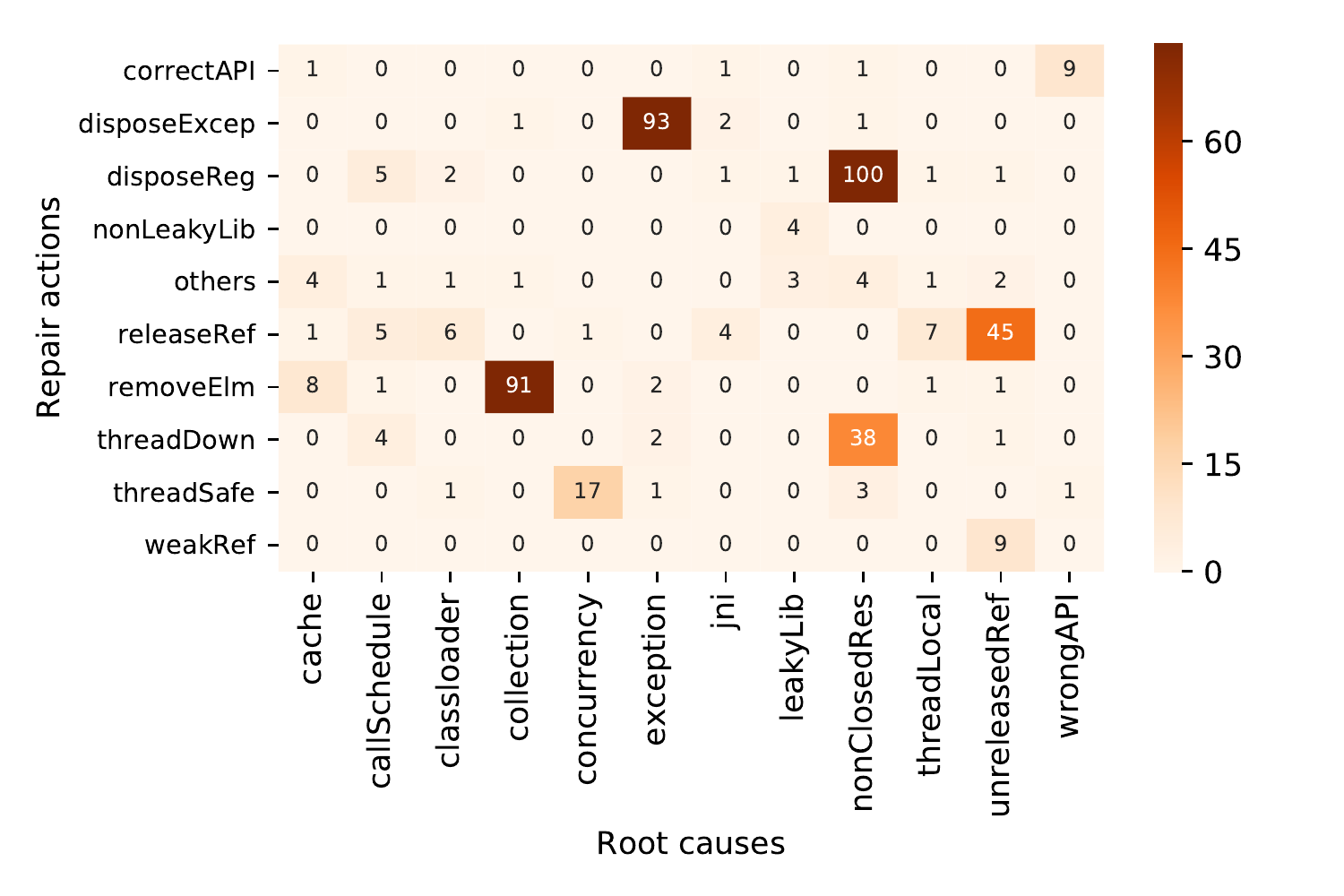}}
\caption{Heatmap of relationship between root causes and repair actions.}
\label{fig:defect-fix-correlation}
\end{figure}
\subRQ{Results.}
In following, we show the results of our analysis on the repair patches in three sub-questions. First, we study the frequency of the repair actions. Second, we analyze the mapping between the root causes and the repair actions to find the relationship between these two taxonomies. Finally, we report the common code transformations found in the fix patches. 

\textbf{Finding 1.}~\Cref{tab:repair-actions} lists the common repair actions along with the number of issues corresponding to them. About 93\% of the resource leaks are repaired with three major actions: \textit{disposeReg}, \textit{disposeExcep}, and \textit{threadDown}, while about 73\% of the memory leaks are fixed with two repair actions \textit{releaseRef} and \textit{removeElm}.

\begin{table*}[t]
\caption{Recurring code transformations and examples of code before and after the transformations.}
\label{tab:patterns}
\begin{itemize}
\item \textbf{Code transformation 1}: Conditional disposal of resource.\\
\textbf{Example}: dispose(obj)\( \rightarrow\ \)If (obj != null) obj.dispose()\\
\item \textbf{Code transformation 2}: Add disposal method call.\\
\textbf{Example}: 
None \( \rightarrow\ \)obj.dispose()\\
\item \textbf{Code transformation 3}: Add disposal method.\\
\textbf{Example}:None \( \rightarrow\ \)void  dispose(){}\\
\item \textbf{Code transformation 4}: Set obsolete reference to null.\\
\textbf{Example}: None \( \rightarrow\ \)obj=null\\
\item \textbf{Code transformation 5}: Add catch/try-catch block.\\
\textbf{Example}: Type obj = new Type() \( \rightarrow\ \)\\try \{Type obj = new Type()\} exception \{dispose(obj)\}\\
\item \textbf{Code transformation 6}: Add finally/try-finally block\\
\textbf{Example}: Type obj = new Type() \( \rightarrow\ \)\\try \{Type obj = new Type()\} finally \{dispose(obj)\}\\
\item \textbf{Code transformation 7}: Add try-with-resources statement.\\
\textbf{Example}: Type obj = new Type() \( \rightarrow\ \)try \{Type obj = new Type()\} \\
\item \textbf{Code transformation 8}: Change condition expression.\\
\textbf{Example}: If (cond1) obj.dispose() \( \rightarrow\ \) If (cond1 and cond2) obj.dispose()\\
\item \textbf{Code transformation 9}: Change method call parameters.\\
\textbf{Example}: obj.method(x, y) \( \rightarrow\ \)obj.method(x, z) \\
\item \textbf{Code transformation 10}: Change static object to a non-static.\\ 
\textbf{Example}: static Type obj = new Type() \( \rightarrow\ \)Type obj = new Type() \\
\item \textbf{Code transformation 11}: Change to weak reference.\\
\textbf{Example}: new HashMap<key, value>() \( \rightarrow\ \)\\new HashMap<key,WeakReference(value)>() \\
\item \textbf{Code transformation 12}: Replace method call.\\
\textbf{Example}: obj.method1() \( \rightarrow\ \)obj.method2() \\
\item \textbf{Code transformation 13}: Change collection.\\
\textbf{Example}: obj = new <collection1>() \( \rightarrow\ \)obj = new <collection2>() \\
\end{itemize}
\end{table*}


\textbf{Finding 2.}~\Cref{fig:defect-fix-correlation} reveals an almost one-to-one mapping between some root causes and repair actions (e.g., \textit{exception}~\(\rightarrow\)~\textit{disposeExcep}, \textit{collection}~\(\rightarrow\)~\textit{removeElm}, \textit{concurrency}~\(\rightarrow\)~\textit{threadSafe}, \textit{concurrency}~\(\rightarrow\)~\textit{threadSafe}). Leak defects with the root cause type \textit{nonClosedRes} are repaired with repair actions \textit{threadDown} and \textit{disposeReg}. Leak defects with the root cause type \textit{unreleaseRef} are repaired with repair actions \textit{releaseRef} and \textit{weakRef}.



\begin{figure}[t]
\center
\scalebox{0.5}{
\includegraphics{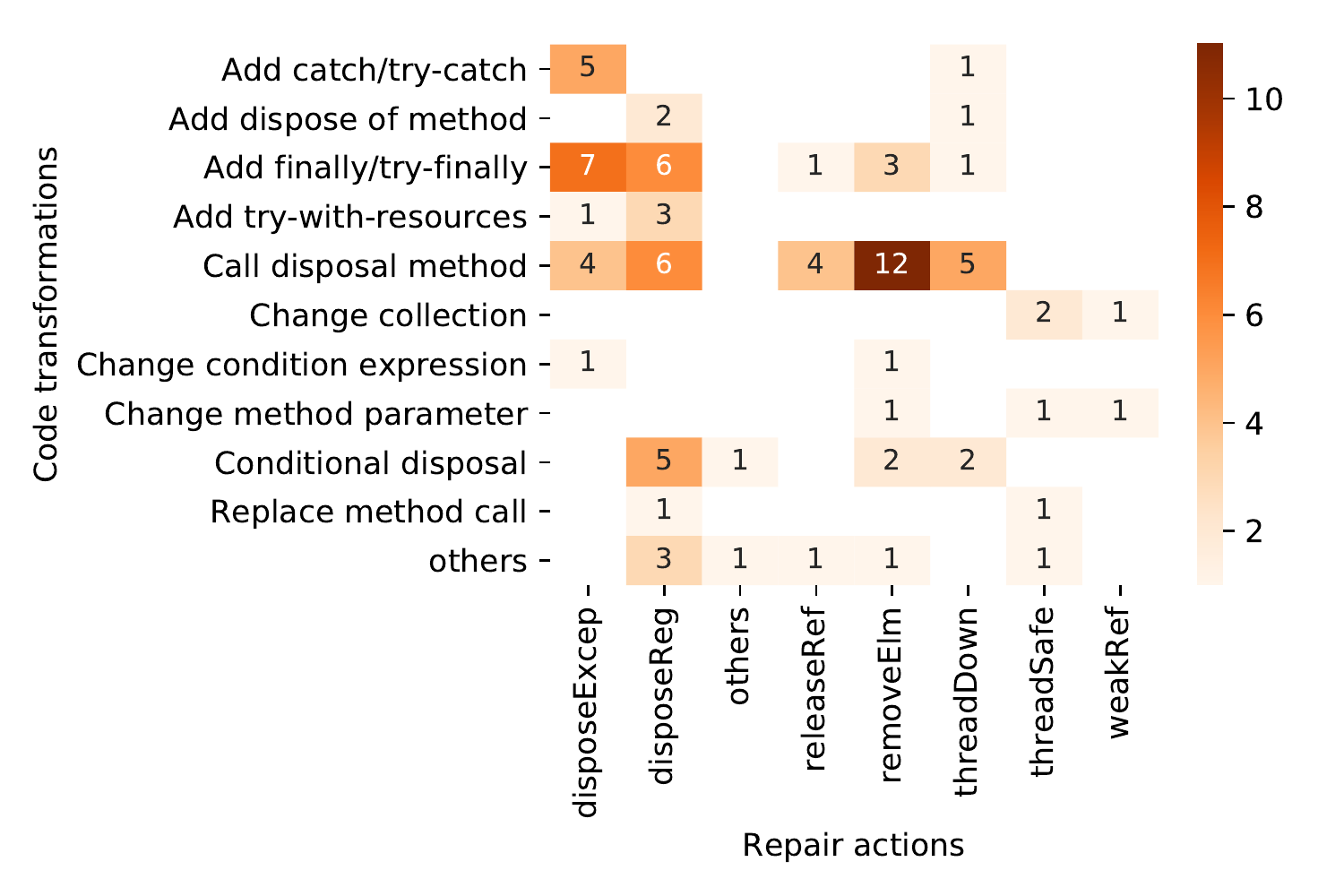}}
\caption{Heatmap of recurring code transformations and single repair actions.}
\label{fig:repair-transformation-heatmap}
\end{figure}


\textbf{Finding 3.} We find 13 recurring patterns in the repair patches. \Cref{tab:patterns} lists the recurring code transformations and the code examples before and after the transformations. Our analysis shows that 88 (out of 491) issues are repaired with a single code transformation. For these issues, we further analyze the quantitative relationship between the repair actions and the most common code transformations.~\Cref{fig:repair-transformation-heatmap} shows the heatmap of the quantitative analysis. For this heatmap, we only consider repair patches with a single code transformation. Code transformation \textit{Add finally/try-finally} is often used in the repair actions \textit{disposeReg} or \textit{disposeExcep}. Code transformation \textit{Add catch/try-catch} is the most used code transformation for repair action \textit{disposeExcep}. We also observe a direct relationship between the repair action \textit{RemoveElm} and the code transformation \textit{Call disposal method}.

This result can encourage the researchers to implement patches for leak-related defects based on template-driven patch generation techniques in the direction of previous work \citep{templateFix,LASEPatchGeneration,27bugPatterns}.

\begin{tcolorbox}
\noindent\textbf{Summary.} Overall, five repair actions are used by developers to repair over 86\% of the issues in our dataset. We found 13 recurring code transformations. 88 of the issues are repaired with a single code transformation. 
\end{tcolorbox}
\subsection{RQ6: \RQVI}
\label{sub:fixComplexity}
\label{RQ600}
\subRQ{Motivation.} This research question addresses the complexity of changes applied to repair the leak-inducing defects. Besides this, we analyze the time to resolve (TTR) for different repair actions. We also compare TTR between leak-related and non-leak-related defects. In this question, we want to find how complex are the repair patches. The results can provide more insights on how complex are the repair patches and how long does it take to repair a leak-inducing defect.


\subRQ{Approach.} To assess the complexity of changes, we compute the code churn and change entropy~\citep{entropyHassan}. 

Code churn is the sum of added and deleted lines in a repair patch. We only consider changes in the code statements and ignore comments or empty lines when calculating the code churn metric. 

We use change entropy to find scatteredness of the changes. Derived from Shannon entropy in information theory, the change entropy measures the complexity of the changes. To measure the change entropy, we use the normalized Shannon entropy~\citep{entropyHassan,ORMEntropy}. It is defined as:
\[H = \frac{- \sum_{i=1}^n p(file_{i}) \cdot log_{e}p(file_{i})}{log_{e}(n)} ,\]
where $n$ is the total number of files in a repair patch and $p(file_{i})$ is defined as the number of lines changed in $file_{i}$ over the total number of lines changed in every file of that repair patch. Change entropy achieves its maximum value when all the files in a repair patch have the same number of modified lines. In contrast, we can achieve a minimum of entropy when only one file has the total number of modified lines. Using the entropy, we can find how complex are the repair patches. The higher the entropy, the more complex the repair patch. 

To asses the time to resolve (TTR) of an issue report, we adopted the methodology used in previous studies~\citep{EmpPerfStat, EmprPerfBugMSR,PerfEmprUnderstand}. We collect two timestamps from each issue report: the time it was created (recorded in the issue tracker), and the time it was resolved (labeled as resolved). For GitHub projects, we use the closed timestamp as the resolved timestamp as it is the only available timestamp in the issue report. For issues with multiple patches, the resolved timestamp is the time of the latest patch being applied to repair the bug. The TTR is the difference between created and resolved timestamps. The reason for using TTR is that the bug trackers used in this study (i.e., Jira and GitHub bug tracker) record no information about the exact amount of coding time needed for fixing a bug.


\begin{figure}[t]
\center
\scalebox{0.5}{
\includegraphics{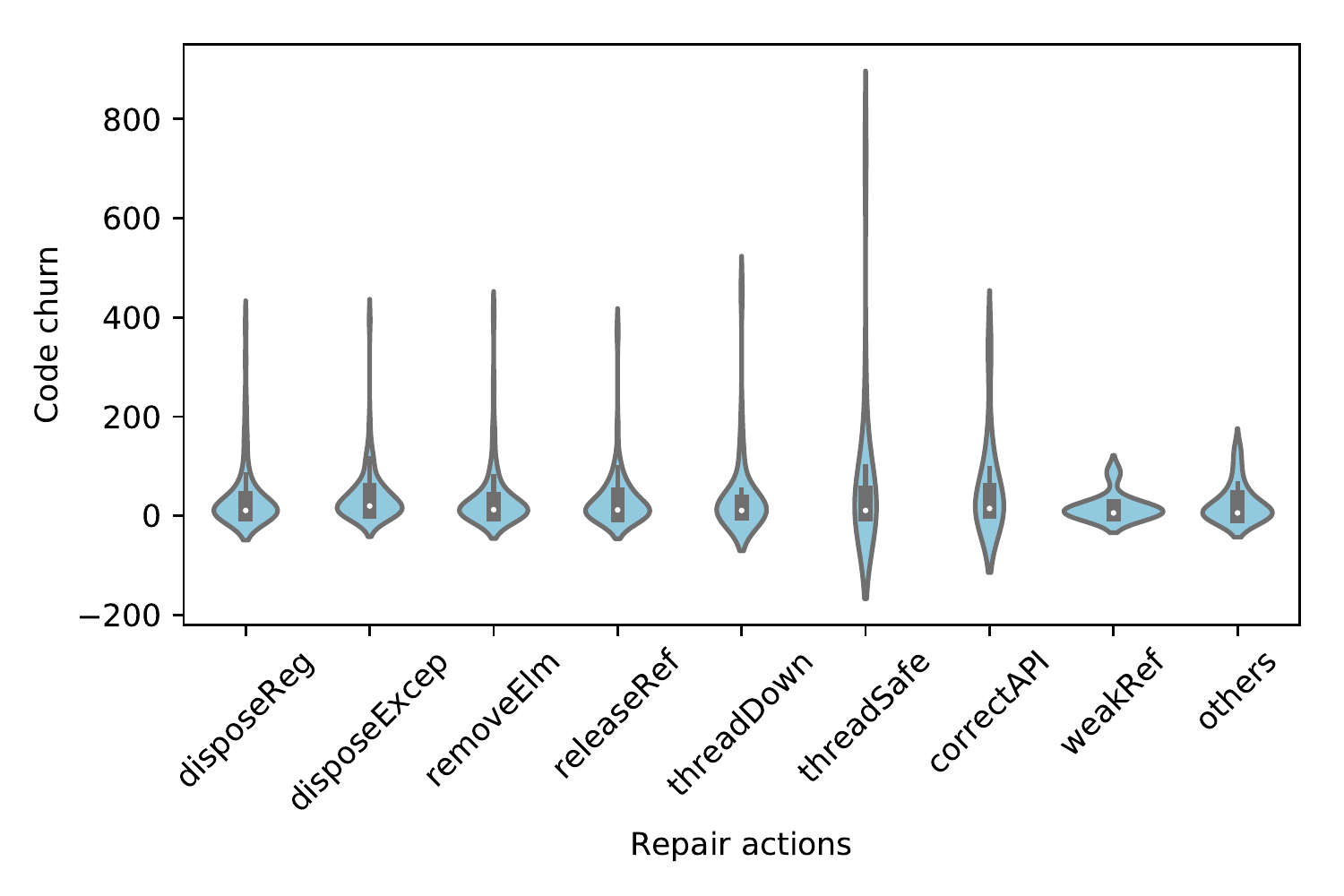}}
\caption{Distribution of code churn per repair action.}
\label{fig:churnFix}
\end{figure}

\subRQ{Results.}
In the following, we show the results of our analysis of the complexity of the repair patches.

\category{Distribution of code churn.}
\Cref{fig:churnFix} shows the box plot of code churn for different repair actions. The line within each box points to the median value of the code churn for that repair action.

\textbf{Finding 1.}
In about all repair actions, the median of code churn is less than 20 lines of code while the repair action \textit{disposeExcp} shows the highest median value.
\begin{figure}[t]
\center
\scalebox{0.45}{
\includegraphics{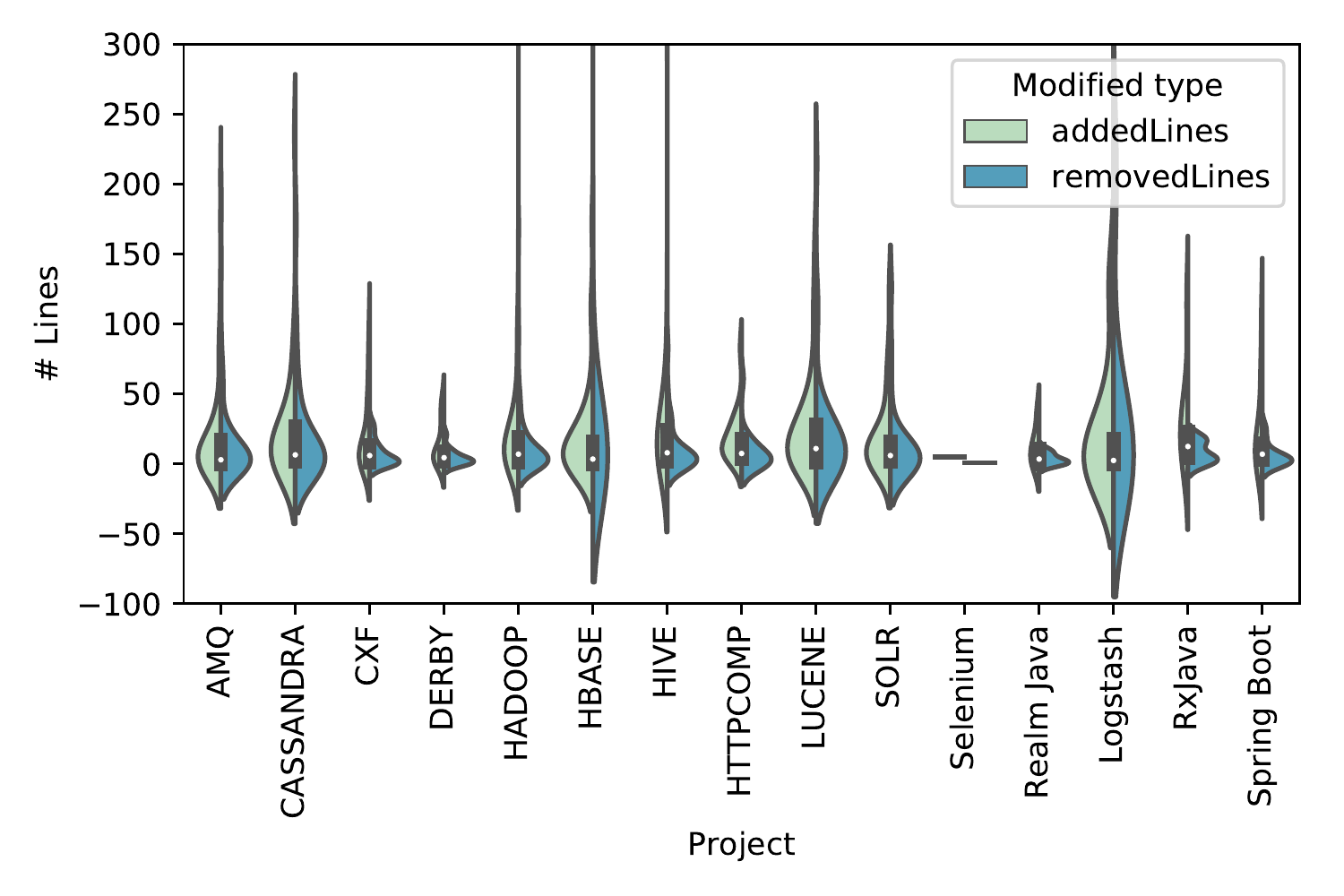}}
\caption{Distribution of number of added and removed lines over the studied projects.}
\label{fig:distAddedRemoved}
\end{figure}

\textbf{Finding 2.}
\Cref{fig:distAddedRemoved} shows the distribution of added and removed lines over studied projects. In all the projects, the median of added lines (29.5 lines) shows a larger value than the removed lines (16.5 lines). Hence, the fault repairing changes often increase the codebase of the applications.

\begin{figure}[t]
\centering
\scalebox{0.4}{
\includegraphics{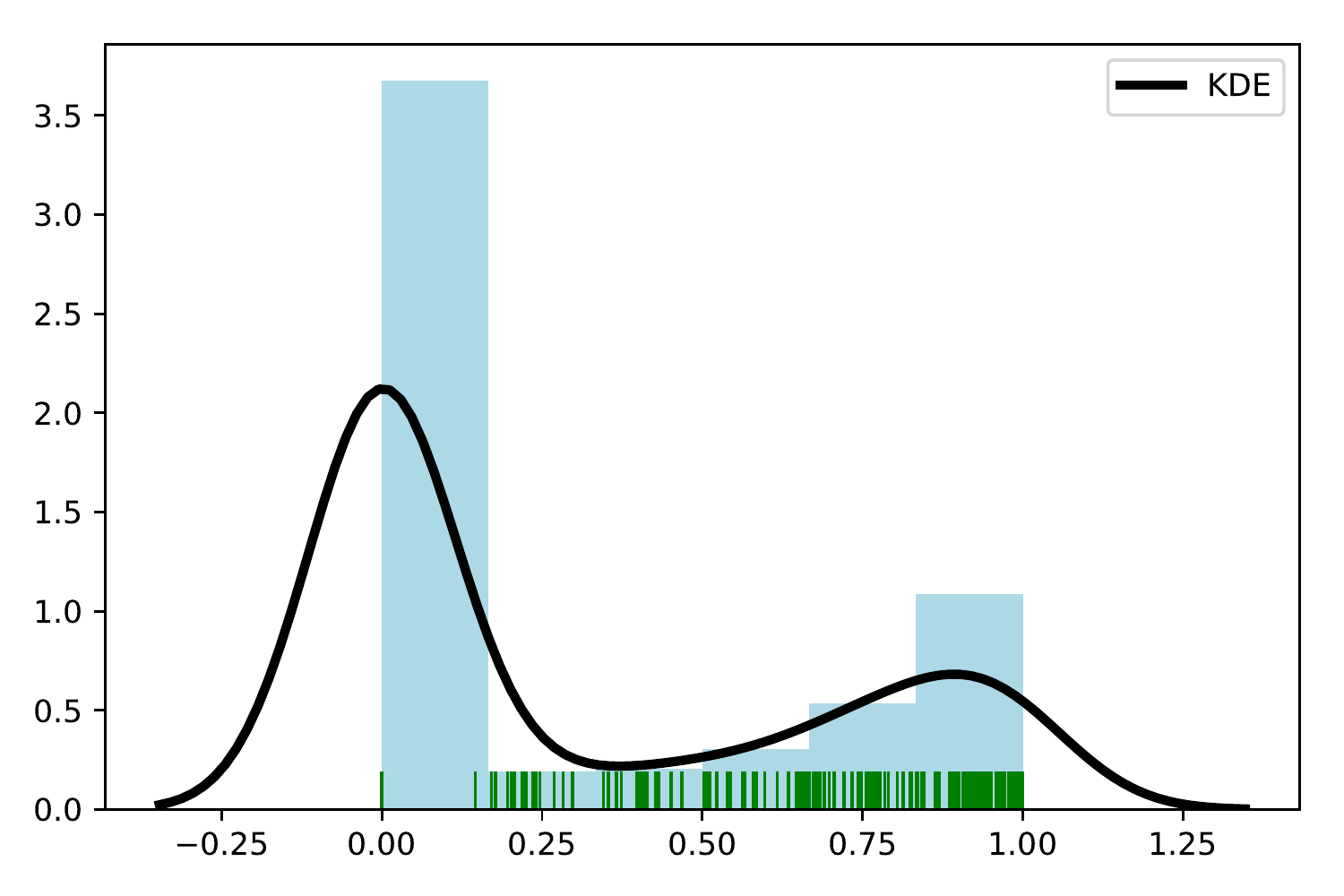}}
\caption{Distribution of change complexity over the repair patches.}
\label{fig:distComplexity}
\end{figure}

\textbf{Finding 3.} \Cref{fig:distComplexity} shows the distribution of change complexity over the repair patches. The distribution appears to be bimodal with the main peak around zero and a lower one around one. The change complexity analysis shows that the changes applied for repairing leak-inducing defects are more concentrated in fewer files and are less scattered. This result can be a useful finding for automated fault localization as it shows the high localization in leak-inducing defects.

\begin{figure}[t]
\center
\scalebox{0.45}{
\includegraphics{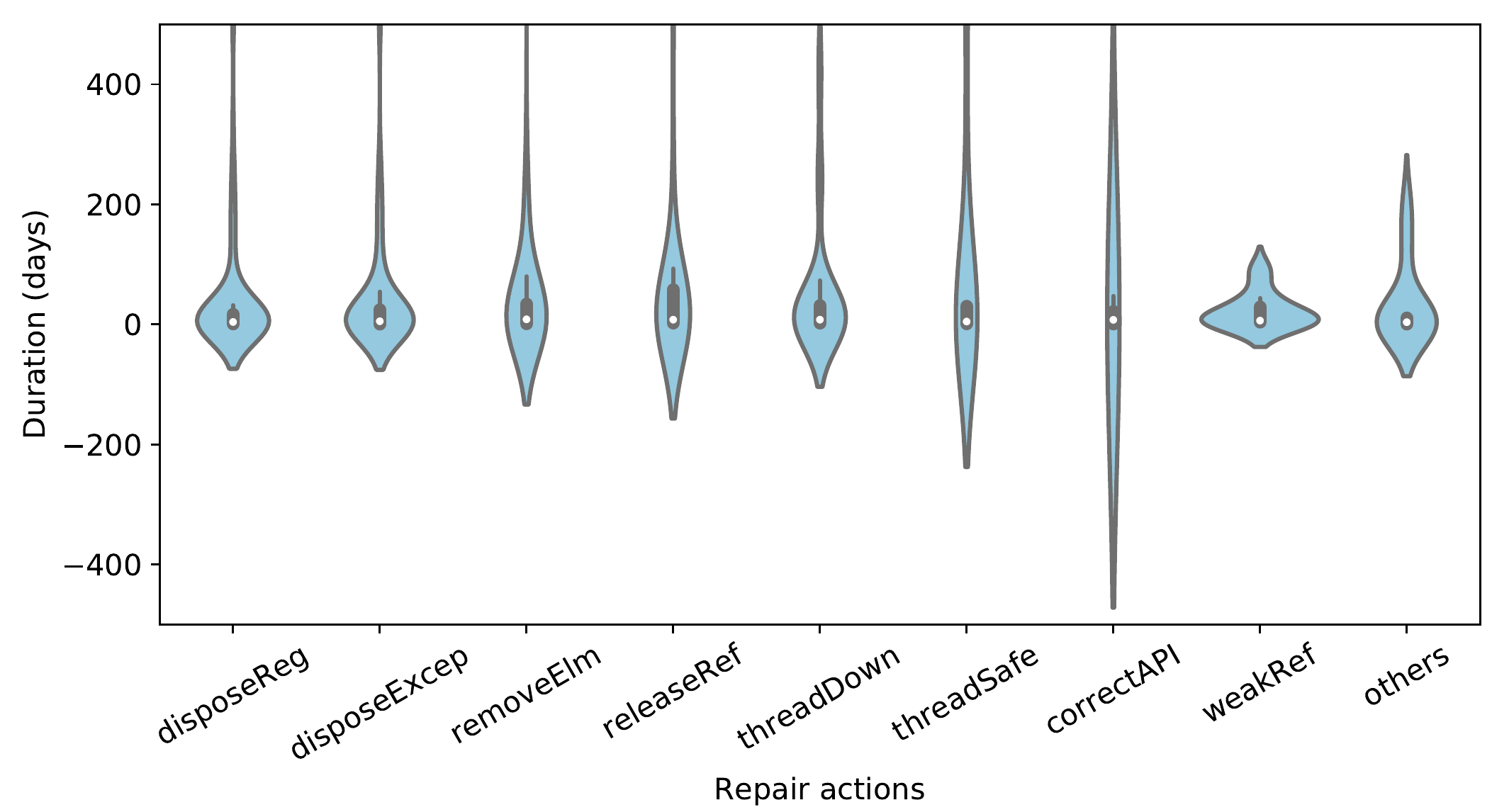}}
\caption{Distribution of time to resolve per repair action.}
\label{fig:durationRepairType}
\end{figure}

\begin{figure}[t]
\center
\scalebox{0.4}{
\includegraphics{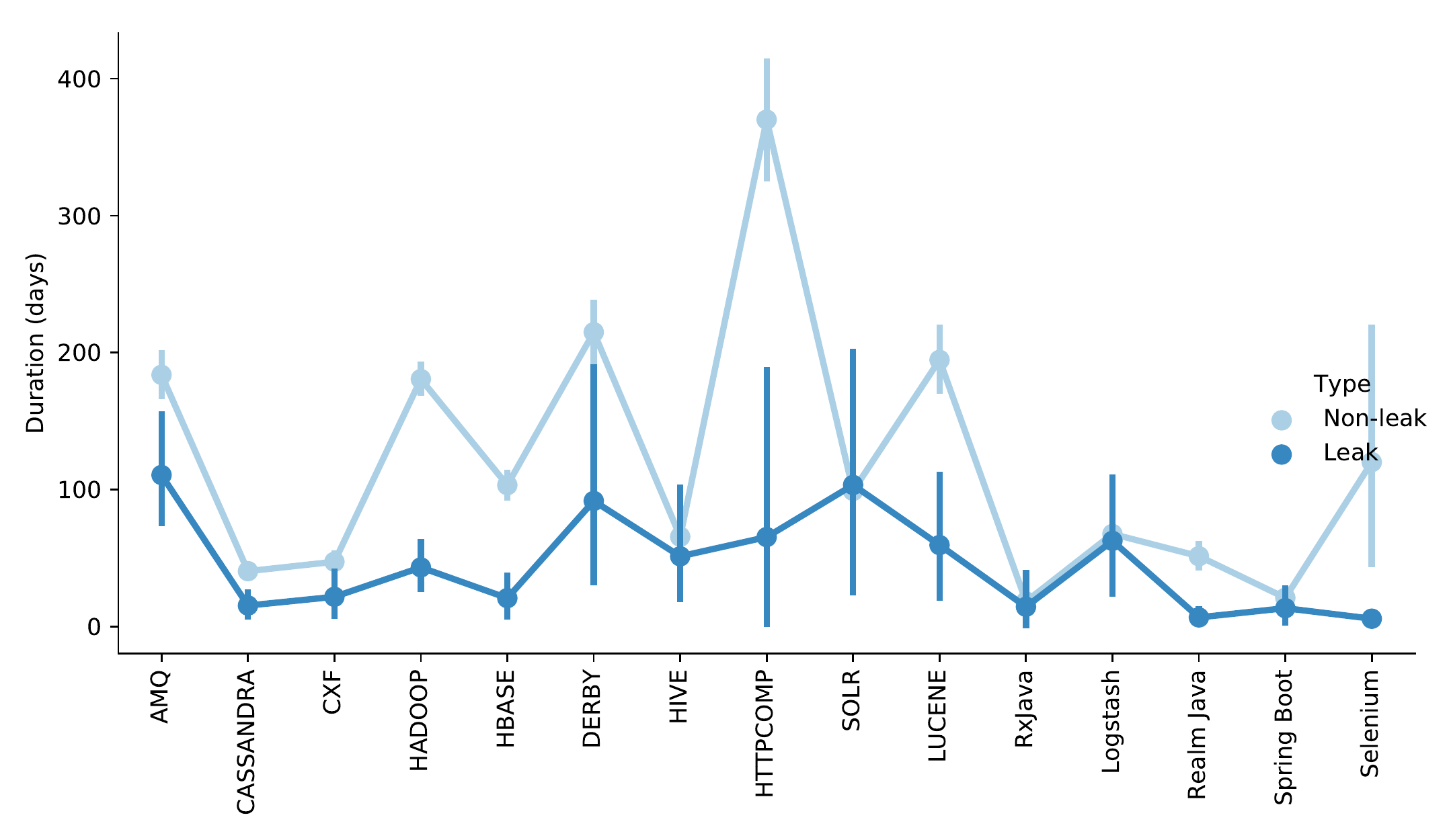}}
\caption{TTR comparison of leak-related and other bugs in studied projects.}
\label{fig:durationComparison}
\end{figure}

\category{Time to resolve (TTR)}
~\Cref{fig:durationRepairType} shows distribution of TTR across repair actions.~\Cref{fig:durationComparison} shows the distribution of the TTR for the leak-related and other defects in the studied projects. To calculate the TTR for other defects, we collect the created and resolved timestamps of all bugs with the resolution ``FIXED'' (except the leak-related defects) from the studied projects in the same time frame that we collected the leak defects.

\textbf{Finding 4.}
On median, about 5.88 days is needed for developers to fix a leak-inducing defect. This time is slightly lower than the TTR for repairing non-leak defects (about 6.04 days). One reason could be that leak-related defects are important for users and developers. The data in our dataset also confirms this. The issue priority in about 84\% of the issues in projects from Apache are labeled as \emph{Blocker}, \emph{Critical}, or \emph{Major} (which are the highest priority levels in the Jira bug tracker). This corroborates with the assumption that leak-inducing defects impose a high negative impact on the performance of the applications, and are highly prioritized by development teams.

\begin{tcolorbox}
\noindent\textbf{Summary.} The change entropy shows that the changes are more concentrated in fewer files and therefore less scattered. The median TTR of the leak-inducing defects is about 5.88 days.
\end{tcolorbox}

\subsection{Other Results}
\label{sec:others}
In this section, we provide other findings found by our study.
\begin{table}[t]
\centering
\caption{The evaluation of Infer static analyzer on a sample of resource leaks from our dataset. Column ``Detected?'' reports whether Infer could detect the defect reported in the respective issue. ``Code-based detection'' refers to source code-based detection. ``Defect'' type and ``Repair'' type are explained in~\Cref{Sec:Defect} and~\Cref{Sec:Fix}, respectively.}
\rowcolors{2}{gray!15}{white}
\begin{tabular}{lllll}
\toprule
\cellcolor{white}\textbf{Issue}&\multicolumn{1}{p{1.2cm}}{\centering \textbf{Detected?}}&\textbf{Detection}&\textbf{Defect}&\textbf{Repair}\\ 
\midrule
\textbf{AMQ-5745}		&	\textbf{\cmark}	&	\textbf{Code-based} 	&	\textbf{nonClosedRes}    &	\textbf{disposeReg}\\
AMQ-6051		&	No	&	Runtime 		&	exception	    &	disposeExcep\\
CASSANDRA-7709	&	No	&	Runtime        &	exception   	&	disposeExcep\\
CASSANDRA-9134	&	No	&	Runtime 		&	nonClosedRes    &	disposeReg\\
\textbf{DERBY-5480}	&	\textbf{\cmark}	&	\textbf{Runtime} 		&	\textbf{nonClosedRes}    &	\textbf{disposeReg}\\
HADOOP-10203	&	No	&	Runtime        &	nonClosedRes    &	disposeReg\\
\textbf{HADOOP-10490}	&	\textbf{\cmark}	&	\textbf{Runtime} 	    &	\textbf{nonClosedRes}	&	\textbf{disposeReg}\\
HADOOP-11014	&	No	&	Code-based 	&	exception		&	disposeExcep\\
HADOOP-11056	&	No	&	Code-based 	&	exception		&	disposeExcep\\
HADOOP-11349	&	No	&	Code-based 	&	exception		&	disposeExcep\\
HADOOP-11368	&	No	&	Runtime 		&	nonClosedRes	&	threadDown\\
HADOOP-11414	&	No	&	Code-based 	&	exception		&	disposeExcep\\
\textbf{HADOOP-9681}	&	\textbf{\cmark}	&	\textbf{Runtime}  &	\textbf{nonClosedRes}	&	\textbf{disposeReg}\\
HBASE-10461		&	No	&	Code-based 	&	exception		&	disposeExcep\\
\textbf{HBASE-10995}		&	\textbf{\cmark}	&	\textbf{Code-based} 	&	\textbf{nonClosedRes}	&	\textbf{disposeReg}\\
HBASE-13601		&	No	&	Runtime 		&	exception		&	disposeExcep\\
\textbf{HBASE-13797}		&	\textbf{\cmark}	&	\textbf{Code-based} 	&	\textbf{nonClosedRes}	&	\textbf{disposeReg}\\
HDFS-1118		&	No	&	Code-based 	&	exception		&	disposeExcep\\
HDFS-1753		&	No	&	Code-based 	&	exception		&	disposeExcep\\
HDFS-5099		&	No	&	Runtime    	&	nonClosedRes	&	disposeReg\\
HDFS-5671		&	No	&	Runtime 		&	exception		&	disposeExcep\\
HDFS-6208		&	No	&	Code-based 	&	nonClosedRes	&	disposeReg\\
\textbf{HDFS-6238}		&	\textbf{\cmark}	&	\textbf{Runtime}        &	\textbf{nonClosedRes}	&	\textbf{disposeReg}\\
HIVE-12250		&	No	&	Runtime 		&	nonClosedRes	&	disposeReg\\
HIVE-12790		&	No	&	Runtime 		&	nonClosedRes	&	disposeReg\\
HIVE-13405		&	No	&	Code-based 	&	exception		&	disposeExcep\\
MAPREDUCE-6528	&	No	&	Runtime 	    &	exception		&	disposeExcep\\
YARN-2484		&	No	&	Code-based 	&	exception		&	disposeExcep\\
\textbf{YARN-2988}		&	\textbf{\cmark}	&	\textbf{Code-based} 	&	\textbf{nonClosedRes}	&	\textbf{disposeReg}\\
YARN-4581		&	No	&	Runtime 		&	exception		&	disposeExcep\\

\bottomrule
\end{tabular}
\label{table:infer}
\end{table}

\subRQ{Efficiency of static analysis tools.} In RQ2 (\Cref{RQ200}), we showed that only in one issue (i.e., CASSANDRA-7709), the resource leak was reported using a static analyzer. There are many static analysis tools for leak detection. They are mostly used for resource leak detection (e.g., FindBugs, Coverity, and Infer). Static analyzers can also be used to detect memory leaks. However, static memory leak detection is imprecise and not scalable for large programs \citep{leakcontainer2013,BloatLeakAnalysis-Xu:2010}. This inefficiency can be largely attributed to the presence of the garbage collector and lack of runtime information. However, one could ask why these tools are not mentioned in the studied issue reports. One reason might be that static analyzers have been employed earlier in the development process, and all leaks detected were fixed.

In our study, we showed that more than half of the studied leaks are resource leaks. It is interesting to study whether static analyzers can detect the studied leak issues. For this purpose, we perform an evaluation of our dataset. We randomly select 30 issues reporting resource leaks from our dataset. As a static analysis tool, we choose Infer which is used by large software organizations\footnote{http://www.fbinfer.com}. We selected Infer because it is an open source tool and can detect resource leaks in Java. The applicability of Infer for resource leak detection is also confirmed in the previous work~\citep{staticRepairInfer}.

\Cref{table:infer} shows the result of our evaluation. From 30 issues, Infer was able to detect the leak defects reported in eight issues. To apply Infer, we first have to build the buggy version of the application in question which contains the leak. After a successful build, Infer produces a file reporting all potential resource leaks. Finally, we investigate whether the file contains the reported leak. We further investigate the eight issues detected by Infer to find the shared characteristics among those issues. In all cases, the leaks occurred in normal execution paths. The analysis shows that Infer was not able to detect leaks triggered in exceptional paths in the sample set. We also observe that developers repaired the leak defects by disposing of the unclosed resources in a~\texttt{try-finally} block. This result can encourage the researcher and tool developers to improve current static analysis tools for leak detection. 

\subRQ{Comparison of common code transformations.} In RQ5 (\Cref{RQ500}), we showed 13 common code transformation found in the studied fix patches. Previous work also reported common repair patterns~\citep{templateFix,R2Fix,27bugPatterns}. PAR~\citep{templateFix} found 10 manual repair patterns for Java. \citep{R2Fix} used 8 patterns (2 of them for repairing memory leaks) to provide patches for bugs in C. \citet{27bugPatterns} introduced 27 automatically extractable repair patterns. 

We compare our 13 patterns with previous work to find which patterns are not reported before.~\Cref{tab:patterns-comparison} shows the result of our evaluation. The result shows that six code transformations were not reported before. We can also observe that ``conditional disposal of resource'' was also used in all studied previous work. The reason why previous work did not report some of the code transformations found by our study may be because they focused on functional bugs, while our study targets leak-related defects. We found that some of the fixes are specific for leak-related defects. For example, \texttt{try}-with-resources is only introduced to avoid potential resource leaks caused by not disposing of closable resources.

\begin{table*}[t]
\centering
\caption{Comparison of common code transformations found in our study with previous work. 27Repairs refers to \citet{27bugPatterns}.}
\begin{tabular}{lrrr}
\toprule
\cellcolor{white}\textbf{Our study}&\textbf{PAR}&\textbf{R2FIX}&\textbf{27Repairs}\\ 
\midrule
Conditional disposal of resource           & \cmark	& \cmark    & \cmark\\
Add disposal method call	                & \xmark	& \cmark    & \xmark\\
Add disposal method			            & \xmark 	& \xmark    & \cmark\\
Set obsolete reference to null		    & \xmark	& \xmark    & \xmark\\
Add catch/try-catch block	            & \xmark	& \xmark    & \cmark\\
Add finally/try-finally block		        & \xmark	& \xmark    & \xmark\\
Add try-with-resources statement		    & \xmark	& \xmark    & \xmark\\
Change condition expression	                & \cmark	& \xmark    & \xmark\\
Change method call parameters	            & \cmark	& \xmark    & \cmark\\
Change static object to a no-static		& \xmark    & \xmark    & \xmark\\
Change to weak reference                    & \xmark 	& \xmark    & \xmark\\
Replace method call                         & \cmark    & \xmark    & \cmark\\
Change collection                         & \xmark 	& \xmark    & \xmark\\
\bottomrule
\end{tabular}
\label{tab:patterns-comparison}
\end{table*}


\section{Implications of the Study} 
\label{implication}
\label{sec:implications}
Based on the findings of our empirical results, we discuss the implications of our study for both researchers and practitioners.

\subRQ{Prevalence of leak types.} Understanding which types of leaks are prevalent in a project can help to avoid and detect leak defects efficiently. The results of~\Cref{RQ100} show that each studied project has a particular dominant leak type. This knowledge can be exploited by prioritizing the most effective detection methods for the dominant leak types. As shown in~\Cref{RQ200}, memory leaks and resource leaks have distinct best practice detection methods which can be used in a software development process. 
Manual code inspection is the dominant detection method for resource leaks. Projects with a large number of resource leaks can benefit from this detection method. One can further improve this by using techniques like code self-review based on the Personal Software Process (PSP)~\citep{PSP2000} with checklist items adapted for detection of resource leaks. For memory leaks, about 63\% of the issues are detected or observed using the runtime information. Projects with a large number of memory leaks should consider the regular usage of the profiling tools. Profiling measures different metrics such as memory or time complexity of a program during its runtime. With this data, programmers can continuously check the resource usage of the program and react faster to the abnormal behavior.

In practical terms, the knowledge of the dominant leak types can be gained via (1) mining distribution of the leak types (or at least the dominant ones) from the bug trackers and repositories, and (2) improving the bug trackers with a labeled classification of the leaked resource.

\subRQ{Good practices.} Good practices can considerably reduce the probability of introducing a leak defect. Such practices can be obtained for example from Java documentation or from existing research work. Here we describe two good practices.

\category{Use try-with-resources on AutoCloseable resources}
Introduced in Java 7, \code{try}-with-resources statement is an efficient method for better management of the closeable resources. It ensures that each resource is closed at the end of the try statement. Our empirical study shows that about 32.4\% of the resource leaks are caused by bad exception handling. The \code{try}-with-resources statement can help to avoid such defects as many current Java applications support Java 7 or higher. 


\category{Prevent having a strong reference from the value object to its key in a HashMap} As opposed to regular references, weak references do not protect the objects from being disposed of by the garbage collector. This property makes them suitable for implementing cache mechanisms through \emph{WeakHashMap}, where the entry will be disposed of as soon as the key becomes unreachable. If the value objects of a HashMap refers to its own key, the programmer should wrap the value as \emph{WeakReference} before putting the value into the map as recommended by the Java documentation\footnote{https://docs.oracle.com/javase/7/docs/api/java/util/WeakHashMap.html}. Otherwise, the key cannot be discarded.



\subRQ{Software testing for leak detection.} Software tests can be used as a lightweight leak detection tool. They are beneficial for decreasing the cost of leak defects by triggering the leaks before the production phase. Our study shows that over 11\% of the leak defects are detected as the result of a failing test (\Cref{Tab:detection-relation}). 
Works like~\citep{perfblower, SCAM2015} corroborate with our results by showing the effectiveness of leak detection via testing. 

\subRQ{Fault localization.} Fault localization is the first step in automated program repair. Defects with high locality can be repaired with low code churn. Our results showed that leak defects are highly localized. First, in 57\% of the issues, only one file was modified. This value increases to 73\% for repairs with changing two files at most. Second, in about 90\% of the issues, only Java files are changed. These findings can encourage researchers to improve and develop techniques for the automated repair of leak defects.

\subRQ{Template-driven patch generation.} Previous works proposed patch-generation techniques based on the templates derived from existing human-written patches~\citep{templateFix, HDRepair}. Work \citep{JSPerformance} showed the existence of common patterns for performance problems in JavaScript. We evaluated the potential of providing template-driven repairs for leak defects through studying repair patches. We found 13 common code transformations used by developers. About 57\% of the issues from patch analysis dataset are repaired by a combination of one or more of these code transformations. These results show the potential of template-driven patch generation techniques for repairing leak-inducing defects.

\section{Threats to Validity} 
\label{sec:threats}
In this section, we discuss the threats to the validity of our study.

\textbf{Construct validity.} The quality of dataset used in our study is a threat to construct validity. First, we used Jira and GitHub bug tracker to collect leak-related defects. This set of defects does not necessarily include all leak defects in the studied applications. Conversely, some investigated defects might never be manifested at runtime. This might be especially the case for issues found by source-code-based detection. Second, to answer research questions, we relied on the information provided in the bug reports as they are the only source of information available. Although the bug reports in the studied projects are often high-quality reports with useful information, it is possible that the reporter provided insufficient information in the report or described the issue incorrectly. However, since we investigate a large number of defects and focus on distributions and their relations, we expect that our findings describe the characteristics of the whole defect population in general.

Second, We used a simple keyword search to find leak-related bugs. Issues that do not contain the keyword ``leak'' can be incorrectly omitted from our data collection process. We searched for other leak-related keywords, but our query yield many false positives. To minimize such threats related to insufficient or skewed sampling of the leak defect population, we used a large set of leak-related bugs (491 issues) from 15 large-scale projects from a variety of application categories and different software repositories. 

Third, we only found one leak-related defect in our dataset in which the leak was detected by a static analyzer. One reason might be that the most leak-related issues are reported on a released version of an application and not during the development phase. It might be the case that the developers already used static analyzers in the development phase to remove the potential leak-related defects in the production environment.

\textbf{Internal validity.} 
Experimenter bias and errors are threats to internal validity. In this study, we heavily used manual analysis for categorization. When generating taxonomies defined in our study, we manually extracted the contents of the issues and used our knowledge to assign a bug to a category. To lower the risk of error in the process of manual categorization, we applied the open coding methodology. Furthermore, the raters discussed any conflicts to reach a consensus in the decision-making meetings. We have spent many hours studying all data related to each defect such as issue title and description, developer discussions, and repair patches. We also computed Cohen's kappa, which is a robust metric for measuring inter-rater agreement. The kappa values ranged from 0.57 to 0.86 that shows a moderate to substantial agreement between the raters. We also make our dataset available online to improve the replicability of our study.

\textbf{External validity.}
Threats to external validity are related to the generalizability of our findings and implications. We collect our dataset from different categories of open source projects. The projects may not be representative for closed source projects. Our results are derived from 15 projects and some findings may not apply to projects written in other languages. 

\section{Related Work} 
\label{relatedWork}
\label{sec:relatedWork}
There is a large body of work in detection, localization, and repairing functional and non-functional bugs. Here we cover work related to our study, grouped in three research directions. 

\textbf{Empirical study.} 
There are many previous works studying characteristics of bugs~\citep{EmprBugFix, PerfEmprUnderstand,EmpPerfStat,JSPerformance}. \citet{EmprBugFix} extracted and analyzed the characteristics of the real bug fixes from six Java projects. Plumbr~\citep{plumbr} reported that over 50\% of mature Java applications contain memory leaks. However, it provided no analysis of the characteristics of the memory leaks. 

Close to our study are work of~\citet{fumio} and \citet{bugStudyOpenSrc} which investigated memory-related bugs. \citet{fumio} investigated five open source Java projects related to cloud computing and found 55 leak-related defects. They showed that in all studied projects leak-related bugs exist with the ratio ranged from 0.4\% to 1.4\% of the total bugs. The majority of 55 leaks were file descriptor leaks comprising of 30\% of the cases.~\citet{bugStudyOpenSrc} studied the characteristics of three open source projects Written in C. They showed that memory-related bugs are one of three main causes of bugs (in addition to concurrency and semantic bugs). They found that 16.7 to 40.0\% of the memory bugs across the studied projects are caused by memory leaks. They also showed that the severity of memory leaks is high as most of them result in a crash. 

Our study differs from the above-mentioned studies. We studied both resource and memory leak-related defects from 15 open source projects. We performed an in-depth analysis of leak defects and their repairs providing taxonomies for leak type, leak detection, fault localization, root-causes, and repair actions. We found that there are common repair patterns for fixing the leak defects. We also evaluated the complexity of the repair patches. Finally, we drawn actionable implications based on our observations and findings. Hence, we believe that our study considerably differs from the previous work in both sizes of the studied dataset and depth of analysis.
 
\textbf{Automated diagnosis of memory and resource leaks.}
Various techniques are proposed by researchers to diagnose memory and resource leaks.

\category{Memory leaks} Static analysis is used to detect memory leaks ~\citep{XieContextLeakStatic2005, HeineCLeak2003,Cheremleak2007,OrlovichLeak2006}. These approaches like other static approaches in fault localization suffer from the lack of scalability and precision. LeakChecker~\citep{LeakChecker2014} tries to decrease the inaccuracy via investigating loops provided by the developer as an oracle for memory leak detection. 

To mitigate the problems of static analysis, other researchers leverages dynamic analysis to detect memory leaks. The major directions of dynamic leak detection are: staleness detection~\citep{SWATLeak2004, BellLeak2006,HoundLeak2009, leakProduction}, growth analysis~\citep{CorkLeak2007, sor, Matias_SysDiffAnalysis2014,perfblower,SCAM2015, VC2012}, analysis of captured
state~\citep{Leakbot2003, leakpoint, leakchaser}, and hybrid approaches~\citep{leakContainerXu2008, OwnershipProfilingLeak2007}.

\citep{leakContainerXu2008} focuses on the memory leak defects related to collections and try to rank the suspicious statements by assigning a leak confidence value based on staleness and memory usage. In our study, we quantitatively show that collections are one of the major root causes of memory leak defects.


Some studies introduced approaches to tolerate the memory leaks by keeping the program in a running state~\citep{ToleratingLeak,OwnershipProfilingLeak2007,LeakPruningBond2009}. They achieve this by reducing performance degradation (e.g., with predicting and reclaiming the leaky objects at runtime).

\category{Resource leaks} Many approaches are used to detect leaks via static analysis including value-flow reachability analysis~\citep{Cheremleak2007}, data-flow analysis~\citep{OrlovichLeak2006}, object ownership analysis~\citep{OwnershipProfilingLeak2007}, loop invariant analysis~\citep{LeakChecker2014}, and automated resource management~\citep{CloserResourceLeak,facade, trackerResourceLeak, weimarErrrorHandling}. They often leverage static analysis techniques to find the unclosed resources on different execution paths. Another research direction is resource leak detection in Android\citep{androidResourceLeak, androidResourceLeakEnergy}.

\textbf{Automated leak repair.}
Recently, automated program repair attracted the attention of researchers. Pioneering work GenProg~\citep{genprog} introduced a patch generation technique based on a genetic search algorithm.~\citet{templateFix} proposed an automated program repair technique based on patterns learned from real patches. It generates correct patches for 27 out of 119 bugs. All the provided fix patterns are simple and one-line statements. Prophet~\citep{Prophet} learns the properties of successful patches to guide finding the appropriate candidates. HDRepair~\citep{HDRepair} leverages information derived from the history of the previous patches of hundreds of Java projects to select the correct patch candidates. All the mentioned techniques differ in defining the search space and the approach to find the accurate patch.

Semantic-based techniques have also been explored~\citep{semfix, angelix}. Angelix~\citep{angelix} is a good example of this category which extracts semantic constraints from the application codebase and generates fixes using program synthesis.

Automated leak repair is still embryonic, and only few works exist in literature~\citep{staticRepairInfer,SafeLeakFixC,valueFlowLeakFixC, R2Fix}.~\citet{staticRepairInfer} purposed Footpatch on top of Infer. Footpatch could generate fixes for resource leaks in C and Java as well as fixes for memory leaks in C.~\citep{SafeLeakFixC,valueFlowLeakFixC} leveraged static, and dynamic analysis to fix memory leaks in C. They analyze the execution paths of each allocation/deallocation and insert a \code{free} when no release is encountered.~\citet{R2Fix} used two repair patterns (\textit{AddFree} and \textit{MvFree}) and provide correct patches for 16 out of 41 memory leaks in C.


\section{Conclusions and Future Work} 
\label{sec:conclusion}
Diagnosis of leak-inducing defects is one of the main challenges for both researchers and practitioners in software development and maintenance. Understanding the characteristics of resource and memory leaks can provide useful information to further improve leak diagnosis techniques. For this purpose, we conducted a detailed empirical study on a large dataset (491 issues from 15 mature projects) to understand how leaks are detected, which defects create them, and which types of repairs exist. Our findings and implications showed that even by simple changes in the quality assurance processes (e.g., code review, testing), the avoidance and diagnosis of leaks could be significantly improved.

In our future work, we will conduct a study on current practices for preventing resource and memory leaks. We will interview Java developers to find out which leak detection tools they use and why they are used. We will also mine the Java codebases to check whether language enhancements such as \texttt{try}-with-resources construct are used, and their impact on the distribution of leak-related defects. We will also evaluate approaches for automated repair of the leak-inducing defects with the focus on template-driven patch generation techniques. We plan to implement a fault injector which simulates the distribution of the leak types and the defect types in real applications. It can serve as a realistic benchmarking tool for the evaluation of methods and tools for leak diagnosis. 

\clearpage

\balance

\end{document}